\documentclass[preprint,review,12pt]{elsarticle}

\usepackage{amssymb}
\usepackage{graphicx}
\usepackage{amsmath}
\usepackage{color}
\usepackage{caption}
\usepackage{subcaption}
\usepackage{lineno}
\usepackage{mathrsfs}

\journal{Computers \& Fluids}

\renewcommand*{\today}{June 13, 2019}

\begin{document}

\begin{frontmatter}



\title{Direct numerical simulation of the multimode narrowband Richtmyer-–Meshkov instability}


\author[label1]{M. Groom\corref{cor1}}
\author[label1]{B. Thornber}

\address[label1]{School of Aerospace, Mechanical and Mechatronic Engineering, The University of Sydney, Sydney, Australia}
\cortext[cor1]{michael.groom@sydney.edu.au}

\begin{abstract}
%\begin{linenumbers}
Early to intermediate time behaviour of the planar Richtmyer--Meshkov instability (RMI) is investigated through direct numerical simulation (DNS) of the evolution of a deterministic interfacial perturbation initiated by a $\textrm{Ma}=1.84$ shock. \textcolor{black}{The model problem is the well studied initial condition from the recent $\theta$-group collaboration [Phys. Fluids. 29 (2017) 105107]. A grid convergence study demonstrates that the Kolmogorov microscales are resolved by the finest grid for the entire duration of the simulation, and that both integral and spectral quantities of interest are converged. Comparisons are made with implicit large eddy simulation (ILES) results from the $\theta$-group collaboration, generated using the same numerical algorithm. The total amount of turbulent kinetic energy (TKE) is shown to be decreased in the DNS compared to the ILES, particularly in the transverse directions, giving rise to a greater level of anisotropy in the flow (70\% vs. 40\% more TKE in the shock parallel direction at the latest time considered). This decrease in transfer of TKE to the transverse components is shown to be due to the viscous suppression of secondary instabilities. Overall the agreement in the large scales between the DNS and ILES is very good and hence the mixing width and growth rate exponent $\theta$ are very similar. There are substantial differences in the small scale behaviour however, with a 38\% difference observed in the minimum values obtained for the mixing fractions $\Theta$ and $\Xi$. Differences in the late time decay of TKE are also observed, with decay rates calculated to be $\tau^{-1.41}$ and $\tau^{-1.25}$ for the DNS and ILES respectively.}
%\end{linenumbers}
\end{abstract}


\end{frontmatter}

%\linenumbers
%\begin{linenumbers}

\section{Introduction}
\label{sec:intro}
The Richtmyer--Meshkov instability (RMI) occurs when an interface separating two materials is impulsively accelerated, such as by a shock wave \cite{Richtmyer1960,Meshkov1969}. The subsequent evolution of the instability is the result of misalignment between the density gradient between the two materials and the pressure gradient driving the impulsive acceleration, referred to as the deposition of baroclinic vorticity, which can be due to a perturbation or inclination of the interface as well as a non-uniformity of the impulse. This deposition leads to a net growth of the interface and the development of secondary shear layer instabilities, which drive the transition to a turbulent mixing layer. Unlike the closely related Rayleigh--Taylor instability, RMI can be induced irrespective of the direction of acceleration, thus both light-heavy and heavy-light configurations are unstable \cite{Brouillette2002}. In both cases the initial growth of the interface is linear and can be described by analytical models. However, as the perturbation amplitude become large with respect to the wavelength the layer growth enters the nonlinear regime and in general numerical simulation is required to calculate the subsequent evolution.  For a comprehensive and up-to-date review of the literature on RMI, the reader is referred to Zhou \cite{Zhou2017a,Zhou2017b}.

The understanding of mixing due to RMI is relevant across a number of scales, ranging from the dynamics of supernovae in astrophysics \cite{Burrows2000}, the mixing of fuel and oxidiser in supersonic combustion \cite{Yang1993}, as well as in inertial confinement fusion (ICF) implosions \cite{Clark2016}. In all of these applications, quantitative experimental data is difficult to obtain, therefore elucidating the underlying physics relies to a considerable extent upon insights gained from numerical simulation. Previous numerical studies of this instability have demonstrated the ability of large eddy simulation (LES) algorithms to predict mixing at late time due to turbulent stirring for Prandtl numbers close to unity \cite{Youngs1994,Hill2006,Thornber2010,Lombardini2012,Tritschler2014,Oggian2015,Liu2016}, with good agreement shown in various integral measures such as width and mixedness across a number of different codes \cite{Thornber2017}.

However, there is still a lack of understanding with regards to the behaviour of the mixing layer during transition, where the turbulence is not fully developed and the energy-containing scales grow under the influence of viscous and diffusive dissipation. Also with regards to ICF, recent simulations have indicated that the capsule hot spot is very viscous due to the high temperatures involved, hence the assumption of turbulent conditions in the hot spot is likely incorrect as small-scale mixing should be viscously damped \cite{Weber2014}. It is also possible that ablator material is spread through the hotspot via molecular diffusion \cite{Clark2016}. \textcolor{black}{This motivates the present paper, which aims to explore the effects of these dissipative processes on the evolution of RMI at early time through direct numerical simulation (DNS) of the governing equations. In particular, the use of DNS is crucial in determining how key flow quantities vary with Reynolds number so as to help identify the conditions that give rise to fully developed turbulence, as well as to provide useful data on how the mixing layer evolves under conditions that inhibit turbulence to some degree.}
	
\textcolor{black}{An additional complication when analysing this transitional behaviour in RMI is that the outer-scale Reynolds number of the layer is not constant but depends on the growth rate (see Section \ref{sec:results}), which in turn depends on the initial conditions \cite{Thornber2010}. Thus it is possible for a mixing layer to become fully turbulent (i.e. the energy-containing scales decouple from the dissipation range) only briefly before decaying to some sub-critical state if the Reynolds number is decreasing in time. This behaviour is not observed in the evolution of other interfacial instabilities such as Kelvin--Helmholtz or Rayleigh--Taylor instabilities, both of which have an outer-scale Reynolds number that increases steadily with time. Therefore the usefulness of DNS in studying RMI is not merely confined to early-time growth and transitional behaviour but also late-time decay and slowly developing mixing layers.}

Previous published direct numerical simulations of RMI include a study by Olson and Greenough \cite{Olson2014}, as well as the studies of Tritschler \textit{et al.} \cite{Tritschler2013,Tritschler2014pre}. In \cite{Olson2014}, single-shock RMI in air and sulphur hexafluoride (SF$_6$) initiated by a Mach 1.18 shock was analysed using two different numerical methods. A maximum grid resolution of $1024\times512^2$ was considered, with an initial perturbation similar to that used in the present study. The initial Reynolds numbers, based on the post-shock velocity and fastest growing wavelength, were 1200 and 7200 in air and SF$_6$ respectively. Using the methodology outlined in \cite{Thornber2017}, and which is also used in the present study (see Section \ref{subsec:setup}), this is equivalent to a Reynolds number of 817 based on the initial mixing width growth rate $\dot{W_0}$ and mean initial wavelength $\overline{\lambda}$. Similarly in \cite{Tritschler2014pre}, RMI initiated by a Mach 1.05/1.2/1.5 shock was simulated, also in air and SF$_6$. A deterministic initial perturbation was used, the maximum grid resolution considered was $1024\times512^2$ and the initial Reynolds number based on $\dot{W_0}$ and $\overline{\lambda}$ was 739.

\textcolor{black}{Given the intention of studying the Reynolds number dependence of transitional behaviour in RMI, it is desirable to use a well defined and well understood initial condition. It is also desirable to start with the simplest possible problem and then gradually introduce additional levels of complexity once the previous level has been well understood. The aim of this paper is therefore twofold. Firstly, it presents a thorough assessment of grid convergence for this specific initial condition. Given that this case was formulated to help understand and compare the performance of individual algorithms, the results presented here should serve as a useful guide in estimating the grid resolution requirements in other DNS studies of RMI, either with different numerical algorithms, or different initial conditions. Secondly, comparisons will be drawn, where appropriate, between the current DNS and previous ILES results for this case using the same numerical algorithm. This clearly demonstrates the impact of finite Reynolds number, how far the current results are from the expected high Reynolds number limit, and indicates specifically in which quantities the  effects of viscous dissipation and molecular diffusion are important.}

The layout of the paper is as follows. Section \ref{sec:description}  details the governing equations solved, the numerical algorithm used to solve them, the initial condition, domain size and boundary conditions as well as the diagnostic quantities. Section \ref{sec:convergence} details grid convergence for the present DNS. Results are presented in Section \ref{sec:results} including visualisations, integral mix measures and power spectra and comparisons between DNS and ILES. Finally, the key conclusions are summarised in Section \ref{sec:conclusion}.

\section{Problem Description}
\label{sec:description}
\subsection{Governing equations}
\label{subsec:equations}
The equations solved are the compressible multicomponent Navier-Stokes equations, which govern the behaviour of mixtures of miscible gases. These equations can be written in strong conservation form as follows:
\begin{equation}
\begin{aligned}
& \frac{\partial \rho}{\partial t}+\nabla\cdot(\rho\boldsymbol{u})=0 \\
& \frac{\partial \rho Y_k}{\partial t}+\nabla\cdot(\rho Y_k\boldsymbol{u})=-\nabla\cdot(\boldsymbol{J}_k) \qquad k=1,\ldots,N-1\\
& \frac{\partial \rho \boldsymbol{u}}{\partial t}+\nabla\cdot(\rho \boldsymbol{u}\boldsymbol{u}^T+p\boldsymbol{I})=-\nabla\cdot\boldsymbol{\sigma} \\
& \frac{\partial \rho E}{\partial t}+\nabla\cdot\big(\big[\rho E+p\big]\boldsymbol{u}\big)=-\nabla\cdot(\boldsymbol{\sigma}\cdot\boldsymbol{u}+\boldsymbol{q}_c+\boldsymbol{q}_d) \\
\end{aligned}
\label{eqn:NS}
\end{equation}
In Eqn. (\ref{eqn:NS}), $\rho$ is the mass density, $Y_l$ are species mass fractions, $u_j$ are the mass-weighted velocity components, $p$ is the pressure and $E=e_i+e_k$ is the total energy, where $e_k=\frac{1}{2}\boldsymbol{u}\cdot\boldsymbol{u}$ is the total kinetic energy and the internal energy $e_i$ is determined by the equation of state. All cases presented here use the ideal gas equation of state, given by:
\begin{equation}
p=\rho e_i(\gamma-1)=\rho\mathscr{R}\sum_{k=1}^{N}\frac{Y_k}{M_k}T
\end{equation}
where $T$, $\mathscr{R}$, $M_k$ and $\gamma$ are the temperature, universal gas constant, molecular weight of species $k$ and ratio of specific heats of the mixture respectively. In general, $\gamma$ is given by:
\begin{equation}
\gamma=\frac{\sum Y_kc_{pk}}{\sum Y_kc_{vk}}
\end{equation}
where $c_{pk}$ and $c_{vk}$ are the specific heats for species $k$. It is worth noting that in the general case where $\gamma$ varies with the mixture composition, numerical methods (such as the finite volume method) are unable to preserve pressure equilibrium across a material interface in the inviscid limit \cite{Abgrall1996}. This has important ramifications for DNS as well, as errors in pressure and/or temperature introduced at an interface result in a finer grid resolution being required for convergence, thus the efficiency of the computation can be severely reduced \cite{Thornber2018}. To avoid this issue here, the mixture $\gamma$ is taken to be constant. This is deemed not to affect any conclusions made about the instability, as the late-time behaviour is approximately incompressible (at least for low to moderate shock Mach numbers) and therefore varying $\gamma$ merely varies the initial impulse given to the interface.

For a Newtonian fluid the viscous stress tensor $\boldsymbol{\sigma}$ is given by:
\begin{equation}
\boldsymbol{\sigma}=-\mu\big(\nabla\boldsymbol{u}+(\nabla\boldsymbol{u})^T\big)+\frac{2}{3}\mu\nabla\cdot\boldsymbol{u}\boldsymbol{I}
\end{equation}
where $\mu$ is the dynamic viscosity. The conductive heat flux $q_c$ is given by:
\begin{equation}
\boldsymbol{q}_c=-\kappa\nabla T
\end{equation}
where $\kappa$ is the thermal conductivity. The enthalpy flux, which arises due to changes in internal energy due to mass diffusion \cite{Cook2009}, is given by:
\begin{equation}
\boldsymbol{q}_d=\sum_{k=1}^{N}h_k\boldsymbol{J}_k
\end{equation}
where $h_k$ is the enthalpy of species $k$. For binary mixtures, diffusion velocities are well approximated by Fick's law \cite{Livescu2013}, thus the diffusive mass flux for each species $k$ is given by:
\begin{equation}
\boldsymbol{J}_k=-\rho D_{12}\nabla Y_k
\end{equation}
where $D_{12}$ is the binary diffusion coefficient.

\subsection{Numerical method}
\label{subsec:numerics}
The governing equations are solved using the University of Sydney code \texttt{Flamenco}, which employs a Godunov-type method of lines approach in a structured multiblock framework. Spatial reconstruction of the inviscid terms is performed using a fifth order MUSCL scheme \cite{Kim2005}, augmented by a modification to the reconstruction procedure to ensure the correct scaling of pressure and velocity and therefore reduced numerical dissipation at low Mach number \cite{Thornber2008a,Thornber2008b,Thornber2007d}. The inviscid flux component is then calculated using the HLLC Riemann solver \cite{Toro1994}, while the viscous and diffusive terms are computed using second order central differences. Temporal integration is achieved via a second order TVD Runge--Kutta method \cite{Spiteri2002}. This numerical algorithm has been used extensively for simulations compressible turbulent mixing problems, including single shock and reshocked RMI \cite{Thornber2011a,Thornber2012,Probyn2014,Thornber2015,Zhou2016,Walchli2017}.

\subsection{Multimode Richtmyer--Meshkov instability}
\label{subsec:setup}
The initial condition used for all simulations is nearly identical to that from the $\theta$-group collaboration, a recent study of the late-time behaviour of the Richtmyer--Meshkov instability by eight independent LES algorithms \cite{Thornber2017}. The only modification is to the initial velocity field to ensure that it satisfies the divergence criteria for a diffuse layer between  miscible fluids. 

A material interface separating two fluids is given a narrowband perturbation with length scales ranging from $L/8$ to $L/4$, where $L=2\pi$ m is the cross section of the computational domain. This interface is diffuse, the profile given by an error function with a characteristic initial thickness of $L/32$. \textcolor{black}{The initial perturbation is given by a power spectrum which is constant over the range of initial wavelengths and zero elsewhere. The total standard deviation of the initial perturbation is taken to be $0.1\lambda_\textrm{min}$ to ensure all modes are initially linear.} The individual mode amplitudes and phases are defined by random numbers, generated by a deterministic algorithm such that they are constant for all grid resolutions considered and are identical to those used in the previous ILES study, allowing for a grid refinement study to be performed.  Full details of the derivation of the initial perturbation can be found in Youngs \cite{Youngs2004} and Thornber \textit{et al.} \cite{Thornber2010}. In order to be suitable for DNS, the perturbation has to be modified to include an initial diffusion velocity at the interface \cite{Reckinger2016}. This is done by considering the incompressible limit of a two-fluid mixture \cite{Livescu2013}, which is given by:
\begin{equation}
\nabla\cdot\boldsymbol{u}=-\nabla\cdot\Big(\frac{D_{12}}{\rho}\nabla\rho\Big)
\end{equation}
To improve the quality of the initial condition, three-point Gaussian quadrature is used in each direction to accurately compute the cell averages required by the finite-volume algorithm.

\begin{figure}[h]
	\centering
	\includegraphics[width=0.85\textwidth]{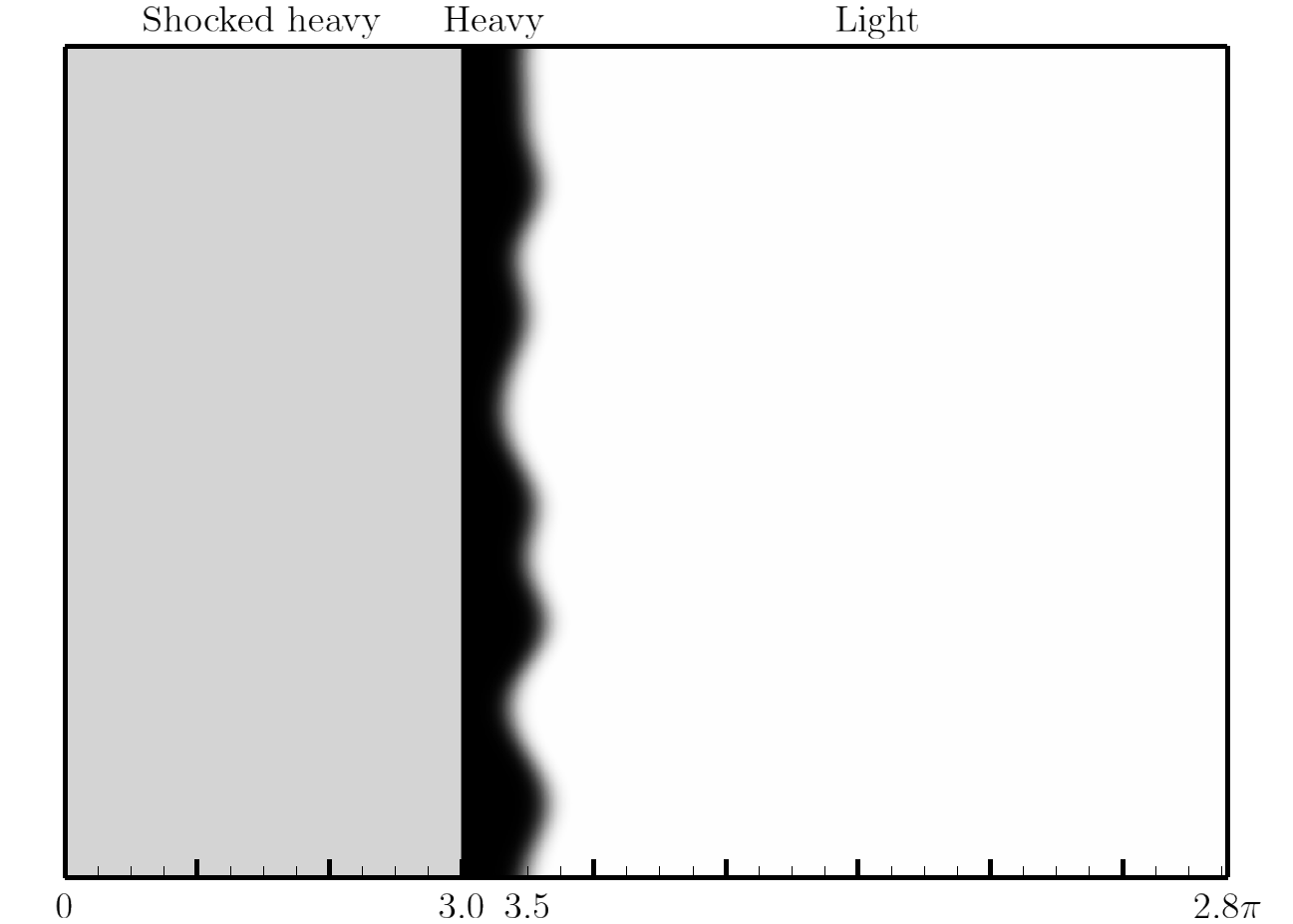}
	\caption{\label{fig:initial} Schematic of the initial condition in the $x$-direction.}%
\end{figure}

The computational domain is Cartesian, with dimensions $x\times y\times z=2.8\pi\times 2\pi\times 2\pi$ m$^3$. Periodic boundary conditions are used in the $y$ and $z$ directions, while in the $x$ direction the outflow boundary conditions are imposed very far from the test section so as to avoid spurious reflections. The initial mean positions of the shock and the interface are $x=3.0$ m and $x=3.5$ m respectively, as shown in Fig. \ref{fig:initial}. The initial pressure of both fluids is 100 kPa, the initial densities of the heavy and light fluids are 3 and 1 kg/m$^3$ respectively and the evolution of the interface is solved in the post-shock frame of reference ($\Delta u=-291.575$ m/s). The shock Mach number is 1.8439, equivalent to a four-fold pressure increase, giving post-shock densities of 5.22 and 1.8 kg/m$^3$ for the heavy and light fluids respectively.

The dynamic viscosity of both fluids is $\mu=0.1$ kg/m/s and is held constant throughout the simulation. The Prandtl and Schmidt numbers of both fluids are $Pr=1.0$ and $Sc=1.0$ and the ratio of specific heats $\gamma$ is 5/3 for both fluids. All non-dimensional quantities presented within this paper are non-dimensionalised by the initial layer growth rate $\dot{W_0}=12.649$ m/s, the mean initial wavelength $\overline{\lambda}=1.0283$ m, the cross -sectional area $A=4\pi^2$ m$^2$ and the mean post-shock density $\overline{\rho^+}=3.51$ kg/m$^3$. This leads to the definition of an initial Reynolds number for the problem, calculated to be $\mathrm{Re}=\dot{W_0}\overline{\lambda}/\overline{\nu}=457$, where $\overline{\nu}=2\mu/(\rho_1+\rho_2)$. For more details on the calculation of $\dot{W_0}$ and $\overline{\lambda}$ see Thornber \textit{et al.} \cite{Thornber2017}.

\subsection{Implicit Large Eddy Simulations}
\textcolor{black}{A brief overview of the previous implicit large eddy simulations performed of this initial condition will now be given here in order to aid comparison with the present set of results. In \cite{Thornber2017}, ILES computations were performed to explore the high Reynolds number limit of key integral quantities in the self-similar regime. Several assumptions are made when justifying the use of ILES as being representative of this high Reynolds number limit. Firstly, it is assumed that there is sufficient separation between the integral length scales and the grid scale such that the growth of the integral length scales are independent of the exact dissipation mechanism. This is addressed in the formulation of the problem, where a sufficient amount of the high wavenumber end of the spectrum is resolved by the grid at all times.}

\textcolor{black}{It is also assumed that the species are intimately mixed at the subgrid level, such that scalar dissipation rates are well represented and are insensitive to the actual values of viscosity and diffusivity. This may be understood as an assumption that the turbulence in the flow is fully developed in the sense of Dimotakis \cite{Dimotakis2000}. Since the effects of physical viscosity and diffusivity on the resolved scales are assumed to be zero, the simulations are nominally inviscid. However, it must be noted that numerical dissipation acts to dissipate kinetic energy and that the equation of state for mixed cells assumes intimate mixing at a sub-grid scale. At early times, when the layer is still non-turbulent and highly corrugated and strained, particular care must be taken in interpreting LES results as the flow is not yet turbulent, thus the thickness of contact surfaces and thus the amount of mixing are impacted by physical diffusion. Once the flow has transitioned to fully developed turbulence however, integral properties of the mixing layer such as width, mixing fractions, turbulent kinetic energy and kinetic energy spectra are considered to be well resolved and an accurate representation of the high Reynolds number limit for this particular flow. At early time it would be expected that quantities dependent on the small scale properties of the layer would differ between the ILES and DNS, yet as the layer transitions to turbulent and becomes well mixed, both LES and DNS are expected to agree. Any disagreement is important as it gives a sense of how far current DNS results are from passing through the mixing transition observed for many other turbulent flows, beyond which point these statistics should become insensitive to the Reynolds number.} 
	
\textcolor{black}{A similar comparison of DNS and ILES results has been made in \cite{Youngs2017} for various flows driven by Rayleigh--Taylor instability, showing that both approaches give very similar results for the degree of molecular mixing at late time. The main conclusions from that paper were that if the high Reynolds number behaviour of the global properties of the mixing zone are of primary interest then ILES results are able to estimate these accurately and with substantially less computational effort. However, if the effects of finite Reynolds number and/or small-scale properties of the mixing zone are desired then DNS (or explicitly modelled LES) is essential.}

 \subsection{Quantities of interest}
\label{subsec:quantities}
To assess the extent of grid convergence, various integral measures are used as well as domain integrated values of some important quantities that are indicative of the fidelity of the numerical simulation at large and small scales. The time dependent integral mixing width $W$ is given by:
\begin{equation}
W(t) =\int_0^{L_x} \langle f_1 \rangle \langle f_2 \rangle \mathrm{d}x
\end{equation}
where $f_k$ is the volume fraction of species $k$ and $\langle\ldots\rangle$ denotes a $y$-$z$ plane average. Similarly the molecular mixing fraction $\Theta$ and mixing parameter $\Xi$ are given by:
\begin{equation}
\Theta(t) =\frac{\int \langle f_1f_2 \rangle \mathrm{d}x}{\int \langle f_1 \rangle \langle f_2 \rangle \mathrm{d}x} \qquad \Xi(t) =\frac{\int \langle \textrm{min}(f_1,f_2) \rangle \mathrm{d}x}{\int \textrm{min}(\langle f_1 \rangle,\langle f_2 \rangle) \mathrm{d}x}
\end{equation}
A third integral mixing measure is also computed, the recently proposed normalised mixed mass \cite{Zhou2016a}:
\begin{equation}
\Psi(t) =\frac{\int \langle \rho Y_1Y_2 \rangle \mathrm{d}x}{\int \langle \rho\rangle \langle Y_1 \rangle \langle Y_2 \rangle \mathrm{d}x}
\end{equation}
At the end of the simulation the integral width is less than 10\% of the domain size, thus the integral measures should be close to statistically converged and not constrained by the choice of domain size \cite{Thornber2016}. The Favre-averaged turbulent kinetic energy is given by:
\begin{equation}
\mathrm{TKE}(t)=\sum_{i}\int \frac{1}{2}\rho u_i''u_i''\mathrm{d}V
\end{equation}
where $u_i''=u_i-\widetilde{u_i}$ and $\widetilde{u_i}=\overline{\rho u_i}/\overline{\rho}$ is a Favre average. \textcolor{black}{The TKE contained in the $i=1$, $i=2$ and $i=3$ directions is denoted by TKX, TKY and TKZ respectively.} Similarly, the domain integrated enstrophy $\Omega$ and scalar dissipation rate $\chi$ are given by:
\begin{equation}
\Omega(t)=\sum_{i}\int \rho \omega_i\omega_i\mathrm{d}V \qquad \chi(t)=\sum_{i}\int D_{12} \frac{\partial Y_1}{\partial x_i}\frac{\partial Y_1}{\partial x_i}\mathrm{d}V
\end{equation}
where $\omega_i$ is the vorticity in direction $i$. \textcolor{black}{The dissipation rate of turbulent kinetic energy is calculated as
\begin{equation}
\epsilon(t)=\sum_{i}\epsilon_i=\sum_{i}\int\overline{\nu}\left(\widetilde{\omega_i^{\prime\prime}\omega_i^{\prime\prime}}+\frac{4}{3}\widetilde{\theta_i^{\prime\prime 2}}\right)\:\mathrm{d}V
\label{eqn:epsilon}
\end{equation}
where  $\theta_i=\partial u_i/\partial x_i$ and $\omega_i^{\prime\prime}$ is the vorticity based on gradients of fluctuating velocities \cite{Chassaing2002}.} \textcolor{black}{Following \cite{Cabot2006}, directional Taylor and Kolmogorov length scales are defined for direction $i$ as
\begin{equation}
	\lambda_{T,i}  =  \left[\frac{\langle u_i^{\prime 2} \rangle}{\langle(\partial u_i^{\prime}/\partial x_i)^2 \rangle}\right]^{1/2} \qquad	\lambda_{K,i}  =  \left(\frac{\langle \nu \rangle ^3}{\langle \epsilon_i\rangle}\right)^{1/4}
	\label{eqn:directional}
\end{equation}
where $\epsilon_i$ is the dissipation rate of TKE in the $i$-th direction, given by Eqn. (\ref{eqn:epsilon}), and plane averages are taken at the mixing layer centre plane, defined as the $x$ location for which $\langle f_1 \rangle=0.5$. Since isotropy is expected in the transverse directions, single transverse Taylor and Kolmogorov scales are defined as $\lambda_{T}=(\lambda_{T,y}+\lambda_{T,z})/2$ and $\lambda_K=(\lambda_{K,y}+\lambda_{K,z})/2$ respectively. Similarly, a transverse Taylor-scale Reynolds number is defined at the mixing layer centre plane as $\textrm{Re}_{T}=(\textrm{Re}_{T,y}+\textrm{Re}_{T,z})/2$, where
\begin{equation}
\textrm{Re}_{T,i} = \frac{\langle u_i^{\prime 2} \rangle}{\displaystyle \langle \nu \rangle \sqrt{\langle(\partial u_i^{\prime}/\partial x_i)^2 \rangle}}
\label{eqn:taylor-Re}
\end{equation}
A Reynolds number based on integral width $W$ is also considered, defined as:
\begin{equation}
\textrm{Re}_W=\frac{W\dot{W}}{\langle \nu \rangle}
\label{eqn:ReW}
\end{equation}}
Finally, radial power spectra of turbulent kinetic energy, enstrophy and scalar dissipation rate are also calculated as follows:
\begin{equation}
E_{\nu}(k)=\hat{\nu}^\dag\hat{\nu} 
\qquad
\begin{cases}
\nu=\sqrt{\rho}u_i'\\      
\nu=\sqrt{\rho}\omega_i\\    
\nu=\sqrt{D_{12}}\frac{\partial Y_1}{\partial x_i}\ \   
\end{cases}
\end{equation}
where $\hat{(\ldots)}$ denotes the Fourier transform and $\hat{(\ldots)}^\dag$ is the complex conjugate of the transform. \textcolor{black}{Power spectra are taken at the $y$-$z$ plane corresponding to the mixing layer centre.}

\section{Grid Convergence}
\label{sec:convergence}
Simulations using the initial condition described in Sec. \ref{subsec:setup} were computed from $\tau=0$ to $\tau=6.15$ using successively refined grid resolutions of $90\times 64^2$, $180\times 128^2$, $360\times 256^2$ and $720\times 512^2$. In order to demonstrate that the results on the $720\times 512^2$ grid were suitably converged at early time, a higher grid resolution of $1440\times 1024^2$ was simulated up to $\tau=6.15$ s. The methods used for determining grid convergence will be discussed in this section. All quantities presented in the following sections are non-dimensionalised as described in Sec. \ref{subsec:setup} and the following colour convention is used to differentiate each grid resolution in the figures: $180\times 128^2$ (green), $360\times 256^2$ (blue), $720\times 512^2$ (black + circles) and $1440\times 1024^2$ (magenta).

\begin{figure}
	\centering
	\begin{subfigure}[t]{\textwidth}
		\includegraphics[width=0.49\textwidth]{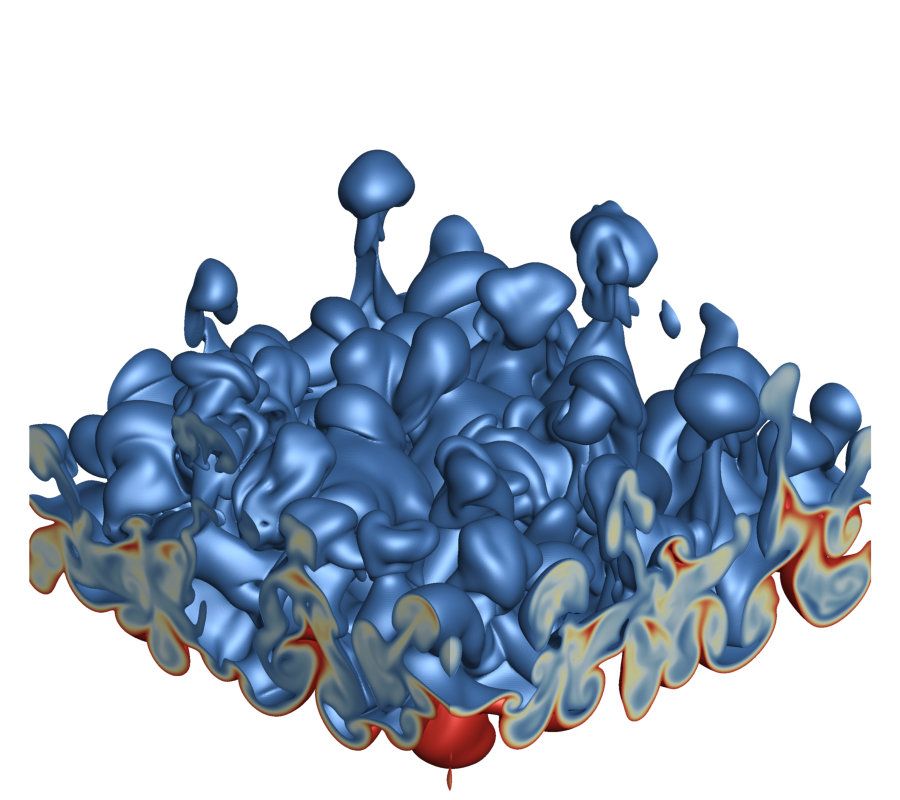}	
		\includegraphics[width=0.49\textwidth]{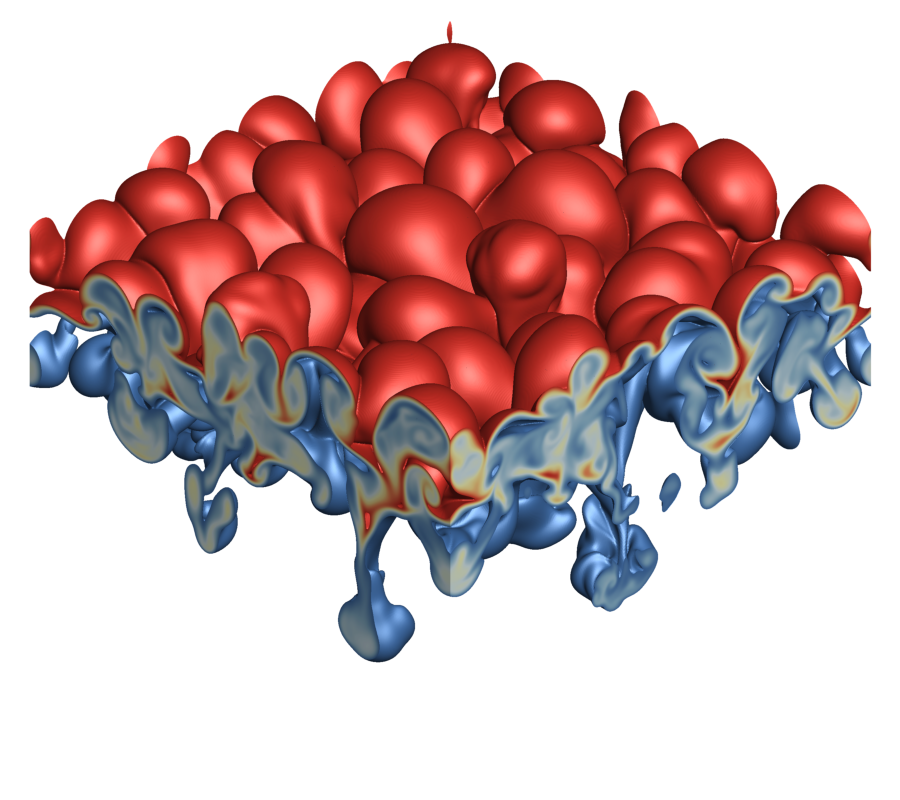}
		\subcaption{$\tau=1.23$}	
	\end{subfigure}
	\hfill	
	\begin{subfigure}[t]{\textwidth}
		\includegraphics[width=0.49\textwidth]{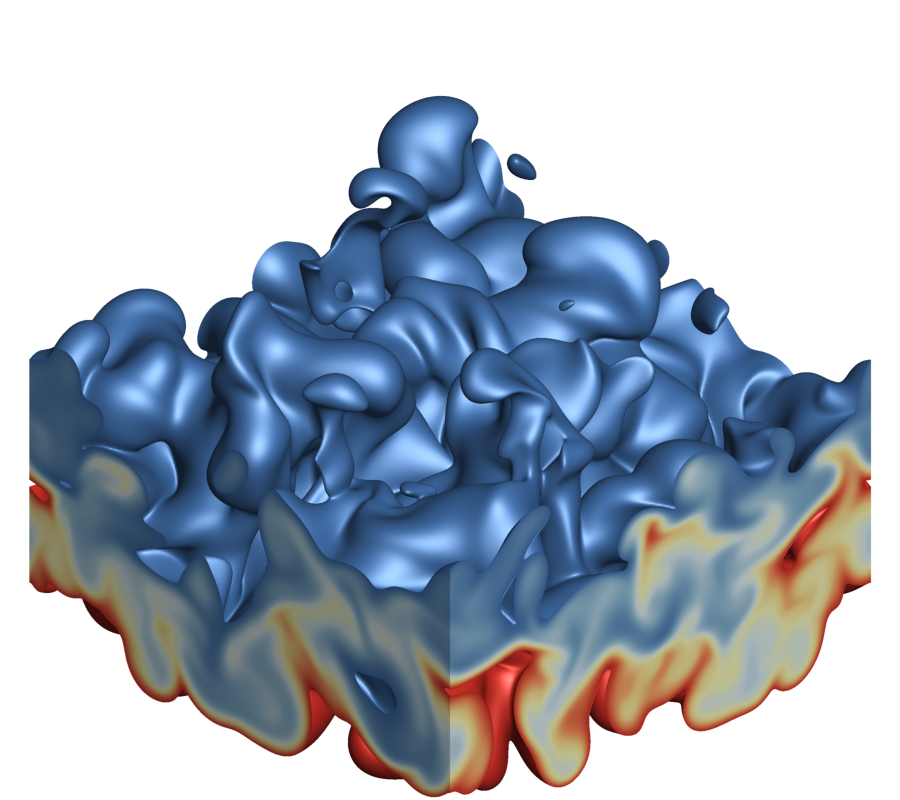}	
		\includegraphics[width=0.49\textwidth]{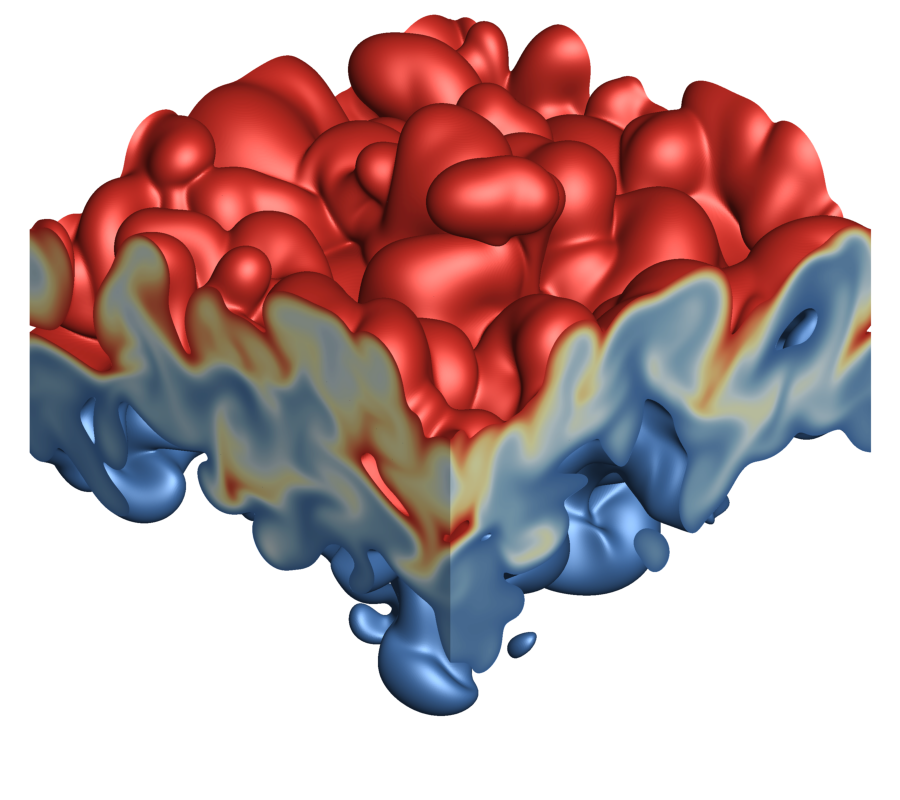}
		\subcaption{$\tau=6.15$}	
	\end{subfigure}
	\hfill
	\caption{\label{fig:isosurface} Contours of volume fraction $f_1$ between the isosurfaces $f_1=0.1$ (blue) and $f_1=0.9$ (red), for the $720\times512^2$ grid.}
\end{figure}

Visualisations of the solution at $\tau=1.23$ and $\tau=6.15$ are given in Fig. \ref{fig:isosurface} showing red bubbles rising into the heavy fluid and blue spikes penetrating into the light fluid. At the end time the mixing layer is still largely laminar with a relatively narrow range of modes, as would be expected for this Reynolds number range, and for the most part the interface has retained a coherent structure. The ability of the numerical algorithm to resolve gradients across the stretched, corrugated layer at early time is the limiting factor in achieving a completely grid independent solution for this case. At later times increased mixing leads to a decrease in mean gradients and hence the resolution requirements are not as severe. 

\begin{figure}
	\centering
	\begin{subfigure}[b]{0.48\textwidth}
		\includegraphics[width=\textwidth]{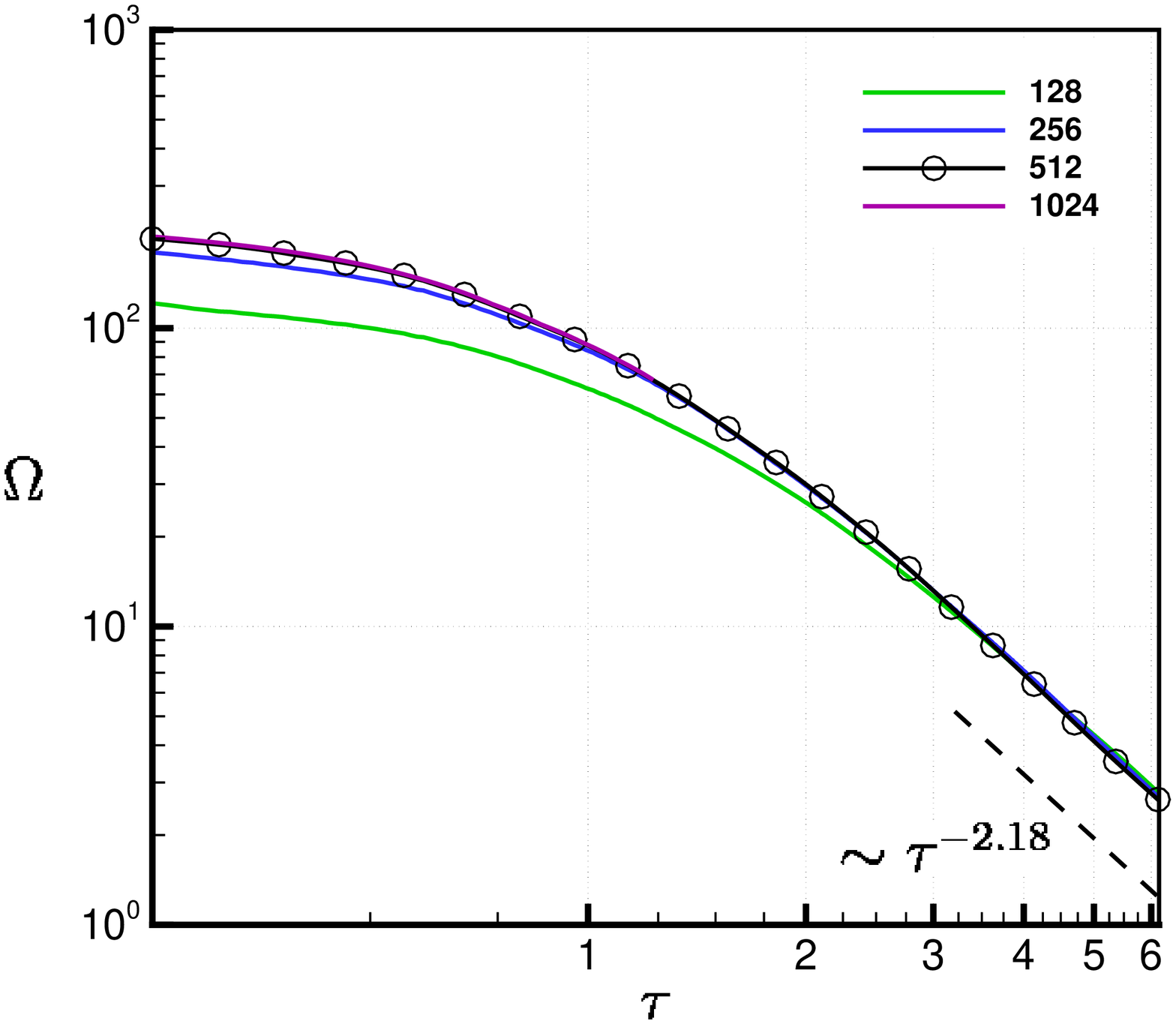}%
		%		\caption{Enstrophy}%
	\end{subfigure}	
	\hfill
	\begin{subfigure}[b]{0.48\textwidth}
		\includegraphics[width=\textwidth]{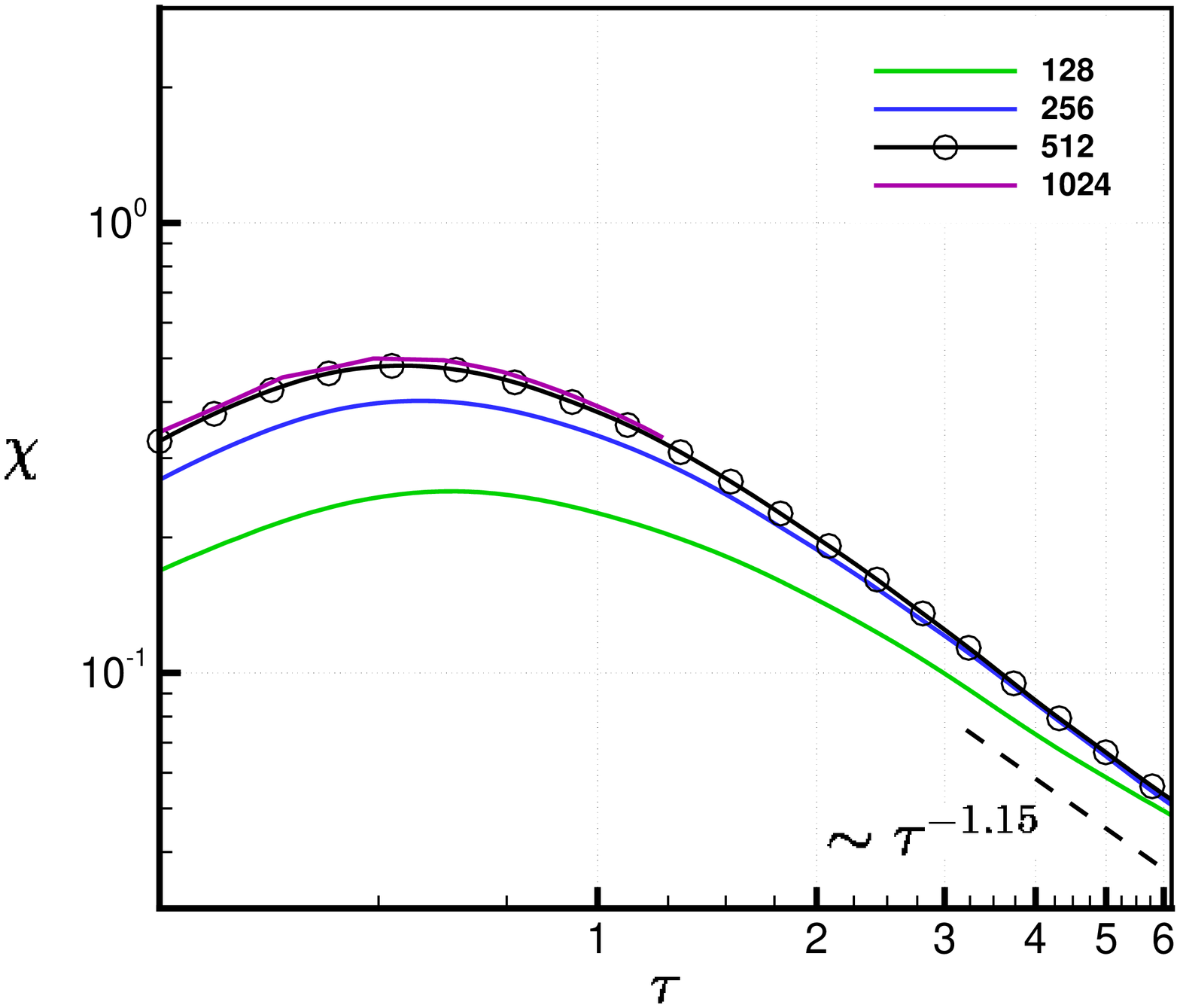}%
		%		\caption{Scalar Dissipation Rate}%
	\end{subfigure}	
	\caption{\label{fig:dissipation} Time histories of enstrophy and scalar dissipation rate.}
\end{figure}

Two domain integrated dissipative measures of the DNS data, enstrophy and scalar dissipation rate, are plotted in Fig. \ref{fig:dissipation}. Grid convergence in these measures is harder to obtain than for lower order statistics \cite{Olson2014}, as this is heavily reliant on the fidelity of the numerical algorithm, in particular how well gradients in the flow field have been captured. Very good agreement is observed for the two finest grid resolutions across all points in time, which gives a good indication that the flow field has been suitably resolved. Fig. \ref{fig:dissipation} also shows that the early time transitional period is the most challenging to fully resolve, which is when the layer is thinnest with respect to the computational grid. 

\begin{figure}
	\centering
	\begin{subfigure}[b]{0.48\textwidth}
		\includegraphics[width=\textwidth]{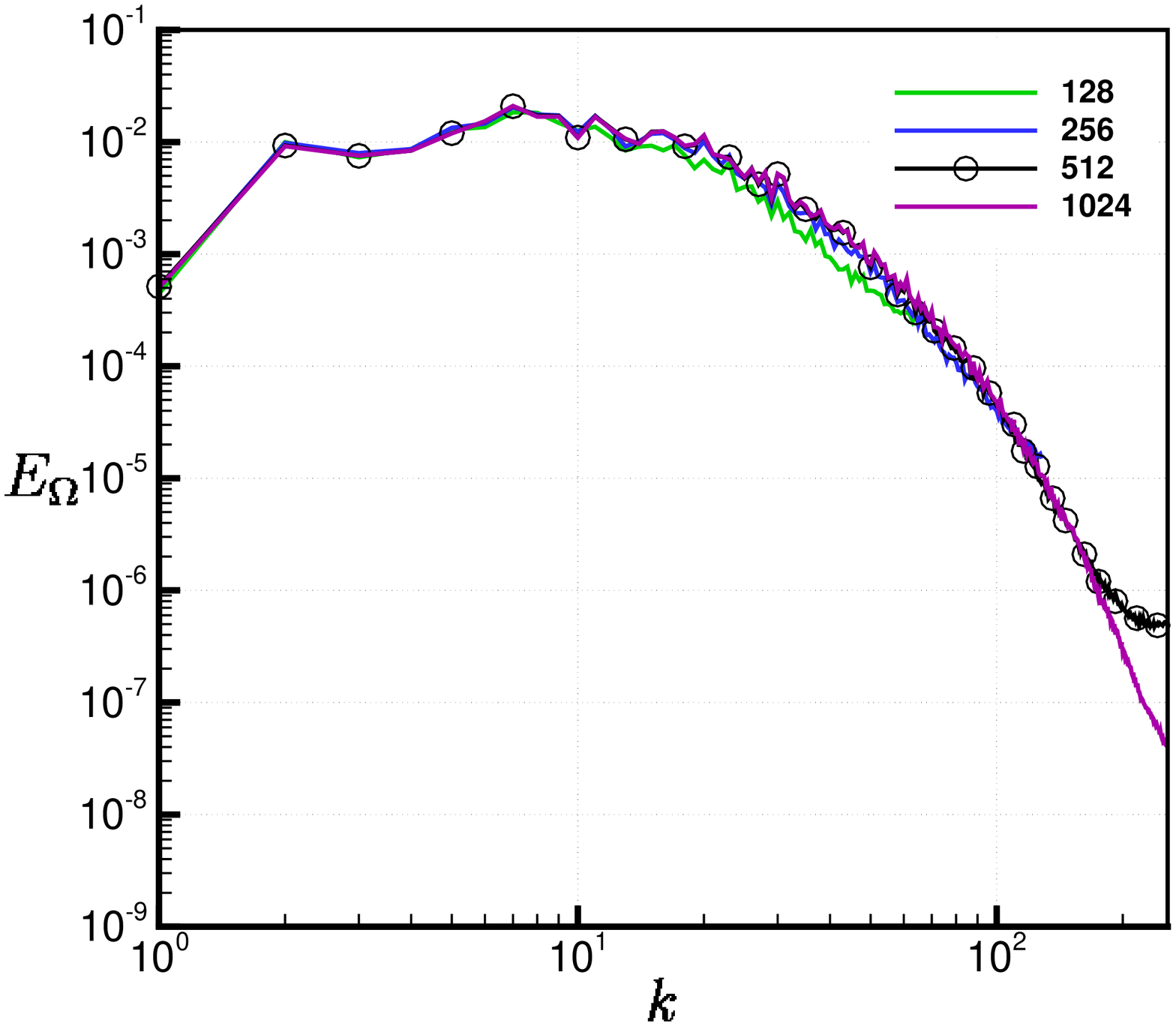}%
		%		\caption{Enstrophy}%
	\end{subfigure}
	\hfill
	\begin{subfigure}[b]{0.48\textwidth}
		\includegraphics[width=\textwidth]{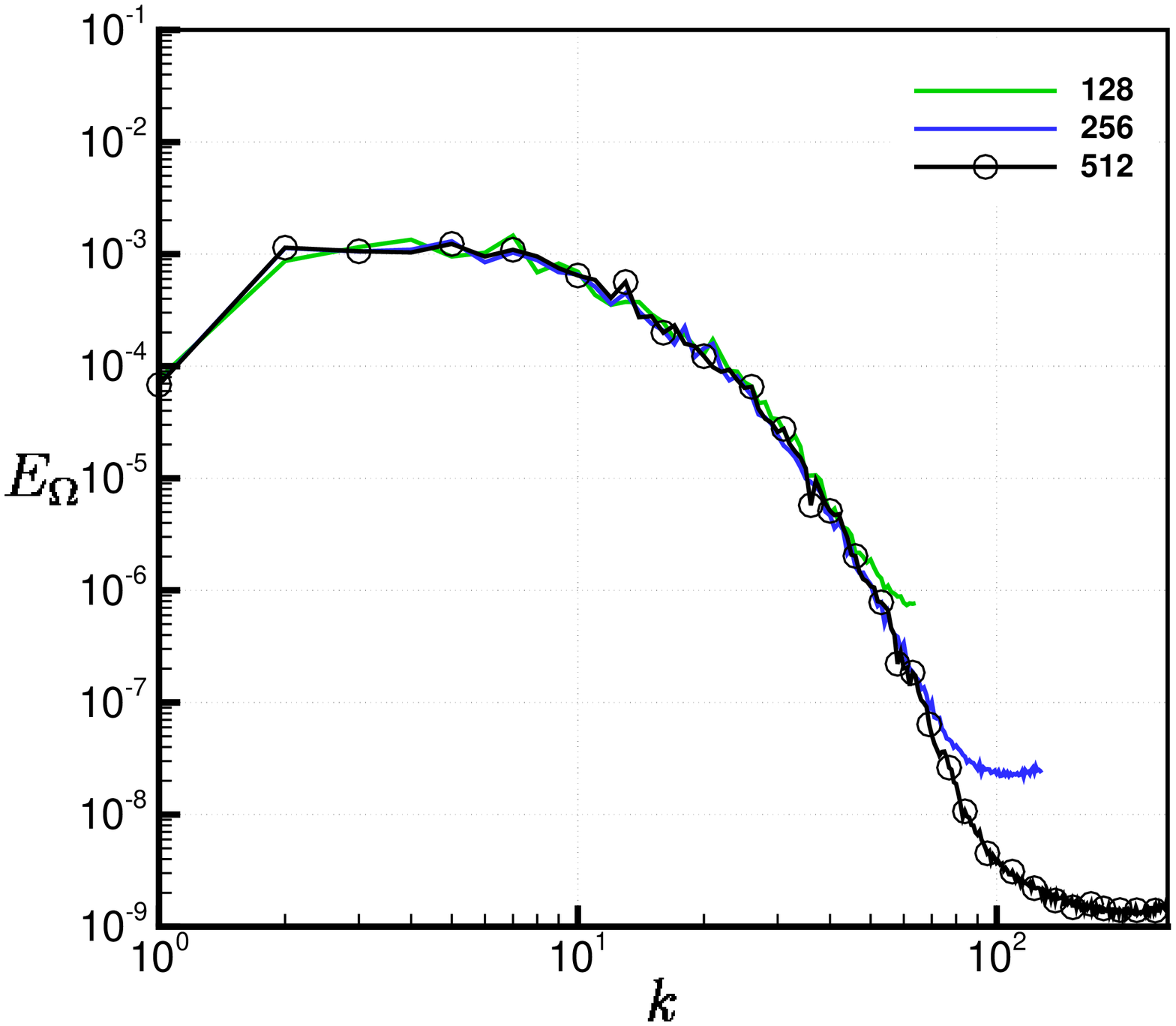}%
		%		\caption{Enstrophy}%
	\end{subfigure}	
	
	\begin{subfigure}[b]{0.48\textwidth}
		\includegraphics[width=\textwidth]{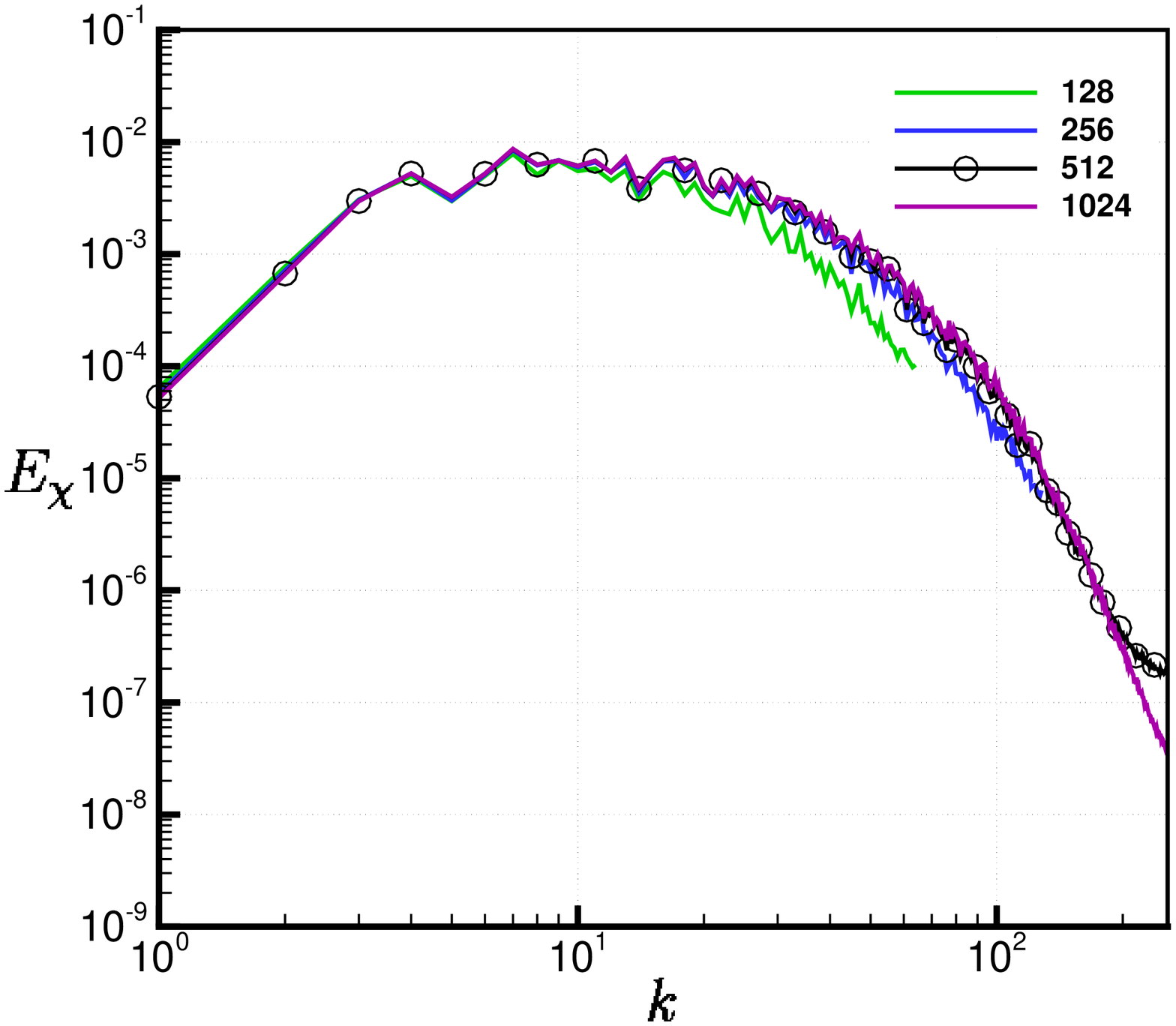}%
		\subcaption{$\tau=1.23$}
	\end{subfigure}	
	\hfill
	\begin{subfigure}[b]{0.48\textwidth}
		\includegraphics[width=\textwidth]{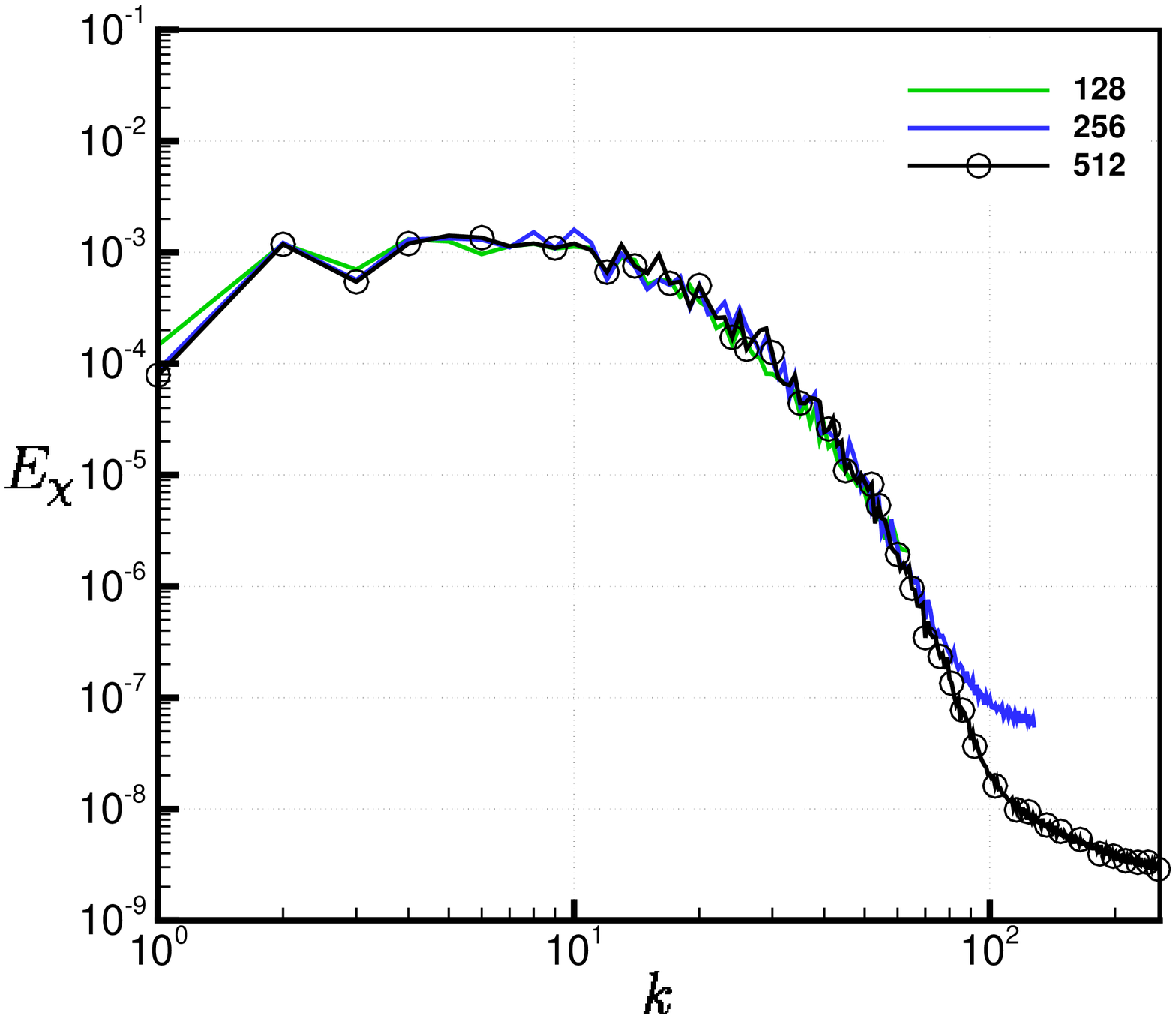}%
		\subcaption{$\tau=6.15$}
	\end{subfigure}	
	\caption{\label{fig:spectra2} Power spectra of enstrophy and scalar dissipation rate.}%
\end{figure}

A more demanding measure is the spectral convergence of these quantities at a given point in time in addition to the domain integrated values. Fig. \ref{fig:spectra2} shows the power spectra of enstrophy and scalar dissipation rate at time$\tau=1.23$ and $\tau=6.15$. Both quantities are grid converged for wavenumbers up to at least $k\approx 100$ throughout the entire simulation, and both display a lack of an inertial range. \textcolor{black}{At the two times considered the Kolmogorov microscale is calculated to be 0.0283 and 0.0624 respectively, equivalent to a wavenumber of $k=222$ and $k=101$, thus the Kolmogorov microscale is less than the Nyquist frequency for both times.} \textcolor{black}{There is a turnup at the high wavenumber end for the higher grid resolutions at late times, where the power is approximately six orders of magnitude lower than the peak. This turnup effect is only observed on grids where the solution is well resolved, being isolated to the far dissipation range of the spectrum (i.e. greater than the Kolmogorov microscale). The most likely source is the reduction of numerical dissipation at low Mach, which is no longer sufficient to remove energy from the very highest modes, however the impact on the spectral range up to the Kolmogorov scale is minimal (it could be filtered out as is commonly done at high wavenumbers when using spectral methods). Comparisons between the spectra for the $720\times 512^2$ and $1440\times 1024^2$ grids at time $\tau=1.23$ (or alternatively the $360\times 256^2$ and $720\times 512^2$ grids at time $\tau=6.15$) show that when the grid resolution is doubled, the dissipation range beyond the Kolmogorov microscale is resolved with no impact on the spectrum at the larger scales. This gives a good indication of the overall fidelity of the numerical method, with discretisation errors being isolated to the highest wavenumbers.}

\begin{figure}
	\centering
	\begin{subfigure}[b]{0.49\textwidth}
		\includegraphics[width=\textwidth]{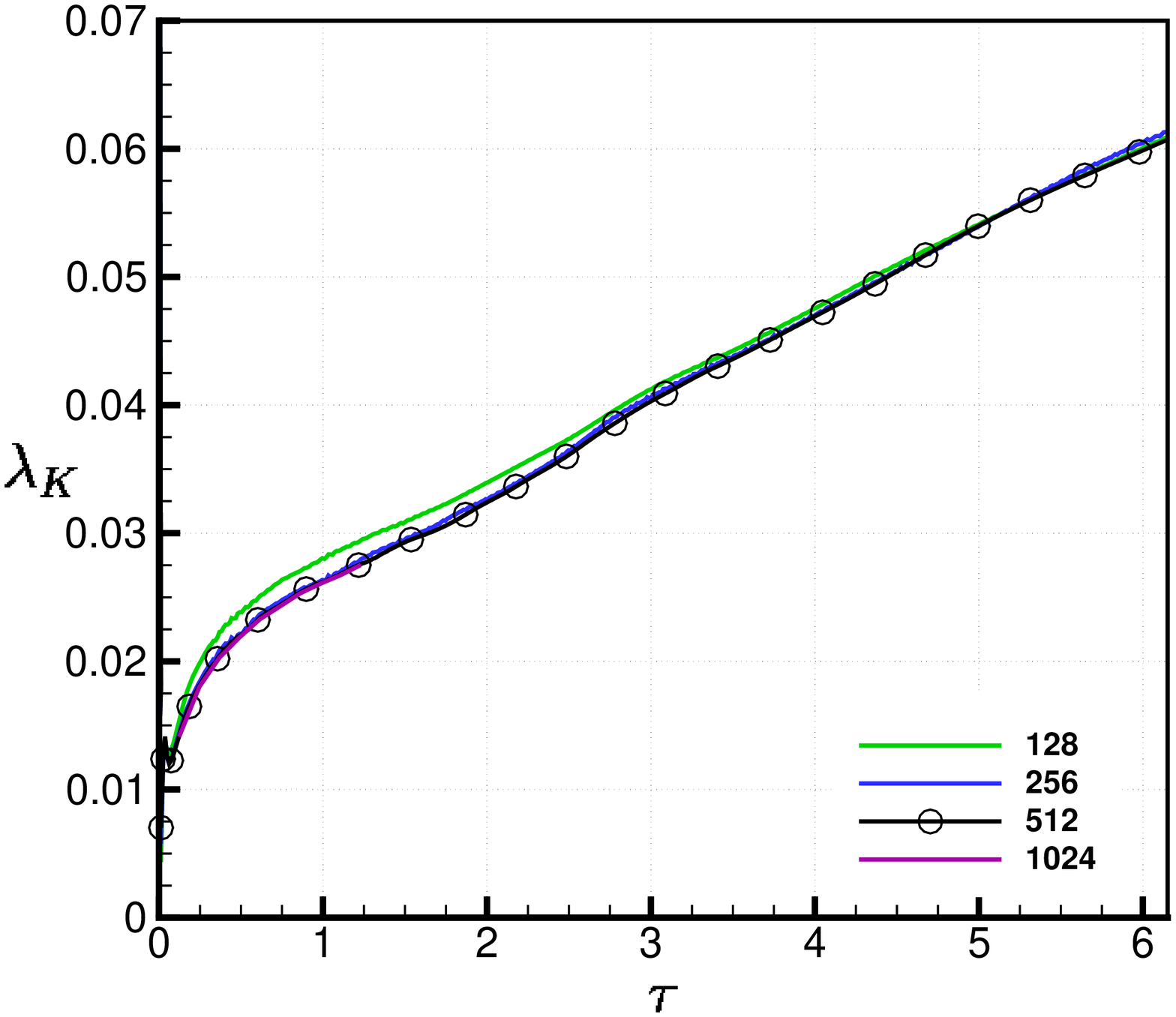}%
		%		\caption{Integral width}%
	\end{subfigure}
	\hfill
	\begin{subfigure}[b]{0.49\textwidth}
		\includegraphics[width=\textwidth]{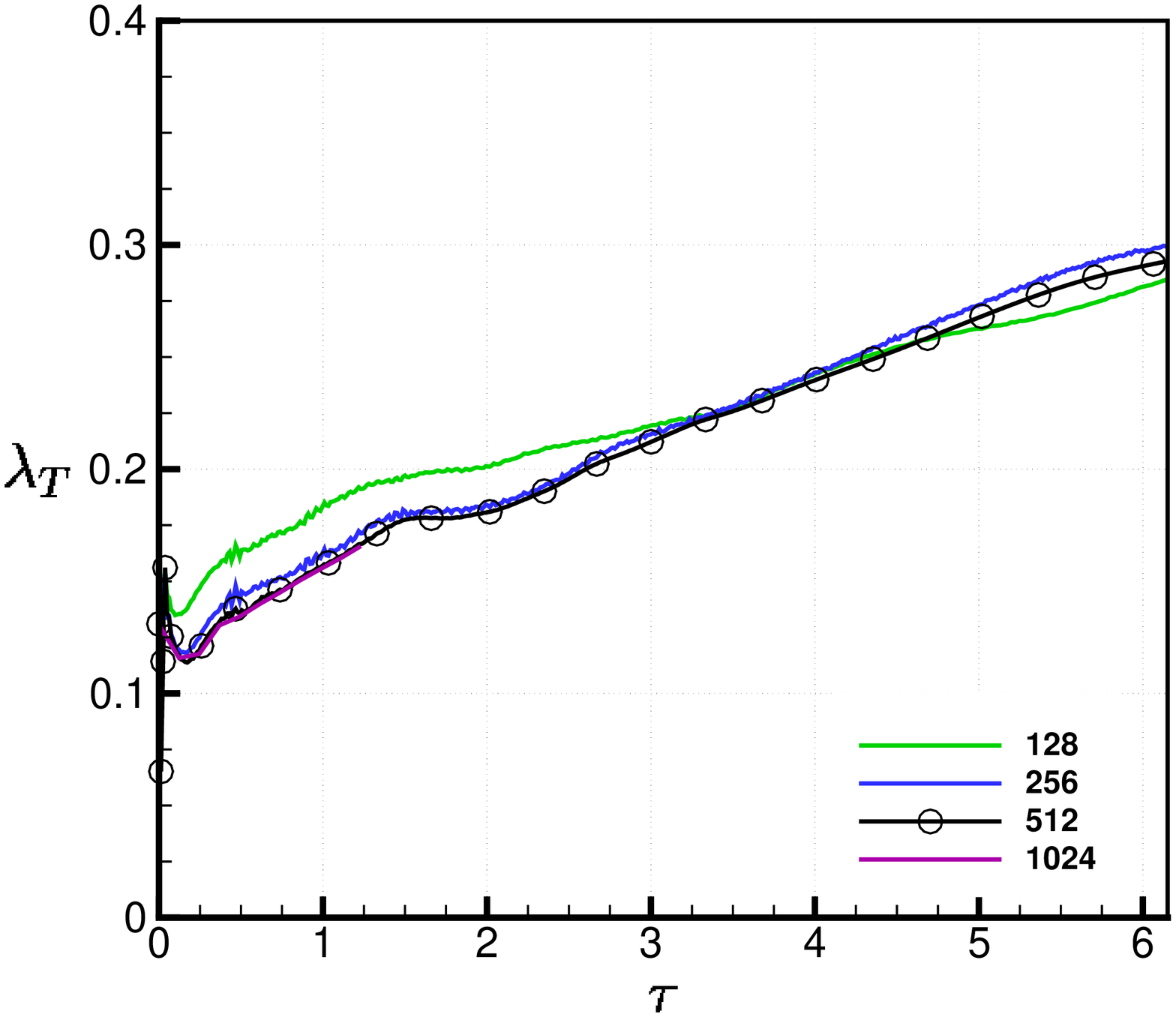}%
		%		\caption{Molecular mixing fraction}%
	\end{subfigure}	
	\caption{\label{fig:lambda} Time histories of Taylor and Kolmogorov microscales, calculated at the mixing layer centre plane.}
\end{figure}

\textcolor{black}{The full temporal evolution of the Kolmogorov microscale, as well as the Taylor microscale, is shown in Fig. \ref{fig:lambda}. Perhaps counter-intuitively, the Taylor microscale is observed to be harder to converge than the Kolmogorov microscale, however both are considered well converged on the $720\times 512^2$ grid across all points in time. With the exception of the very first data point, which is sampled during the interaction of the shock and the interface, $\lambda_K \ge 0.0127$ and is therefore above the finest grid resolution of $\Delta x = 0.0122$ for all subsequent times. Thus it can be concluded that the solution on the $720\times 512^2$ grid qualifies as DNS \cite{Moin1998}. The DNS of Tritschler \textit{et al.} achieved a similar level of resolution, with the Kolmogorov microscale calculated to be between 5 $\mu$m and 92 $\mu$m on a grid spacing of $\Delta x = 19.5$ $\mu$m \cite{Tritschler2014pre}. It should be noted that the validity of the Kolmogorov microscale as being representative of the smallest scales is questionable prior to the flow being fully developed, and as such it is better to focus on the degree to which the results are independent of numerical dissipation (i.e. a grid-converged solution for the quantities of interest for this study).}

\section{Results}
\label{sec:results}
\begin{figure}
	\centering
	\begin{subfigure}[b]{0.49\textwidth}
		\includegraphics[width=\textwidth]{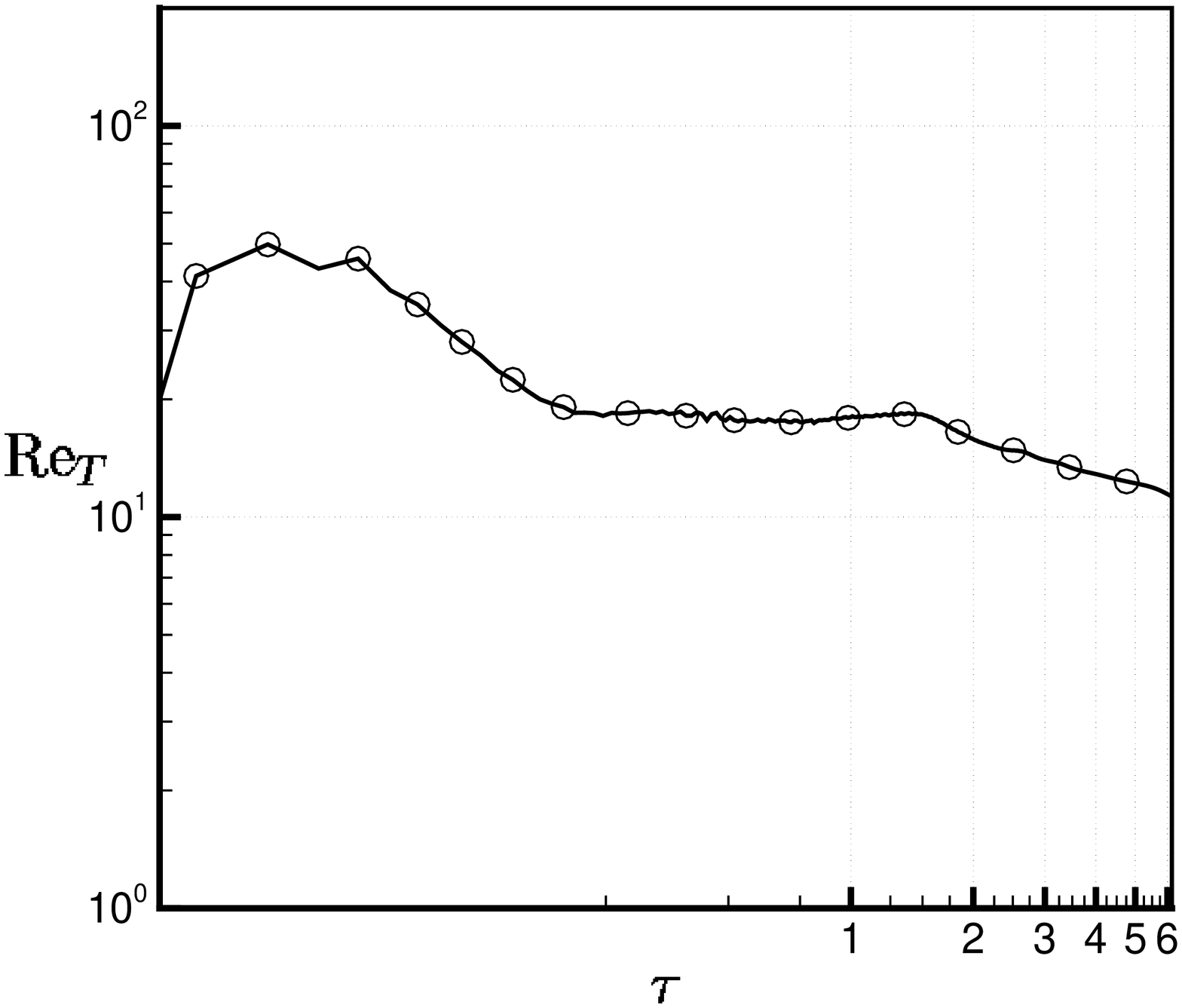}%
		%		\caption{Integral width}%
	\end{subfigure}
	\hfill
	\begin{subfigure}[b]{0.49\textwidth}
		\includegraphics[width=\textwidth]{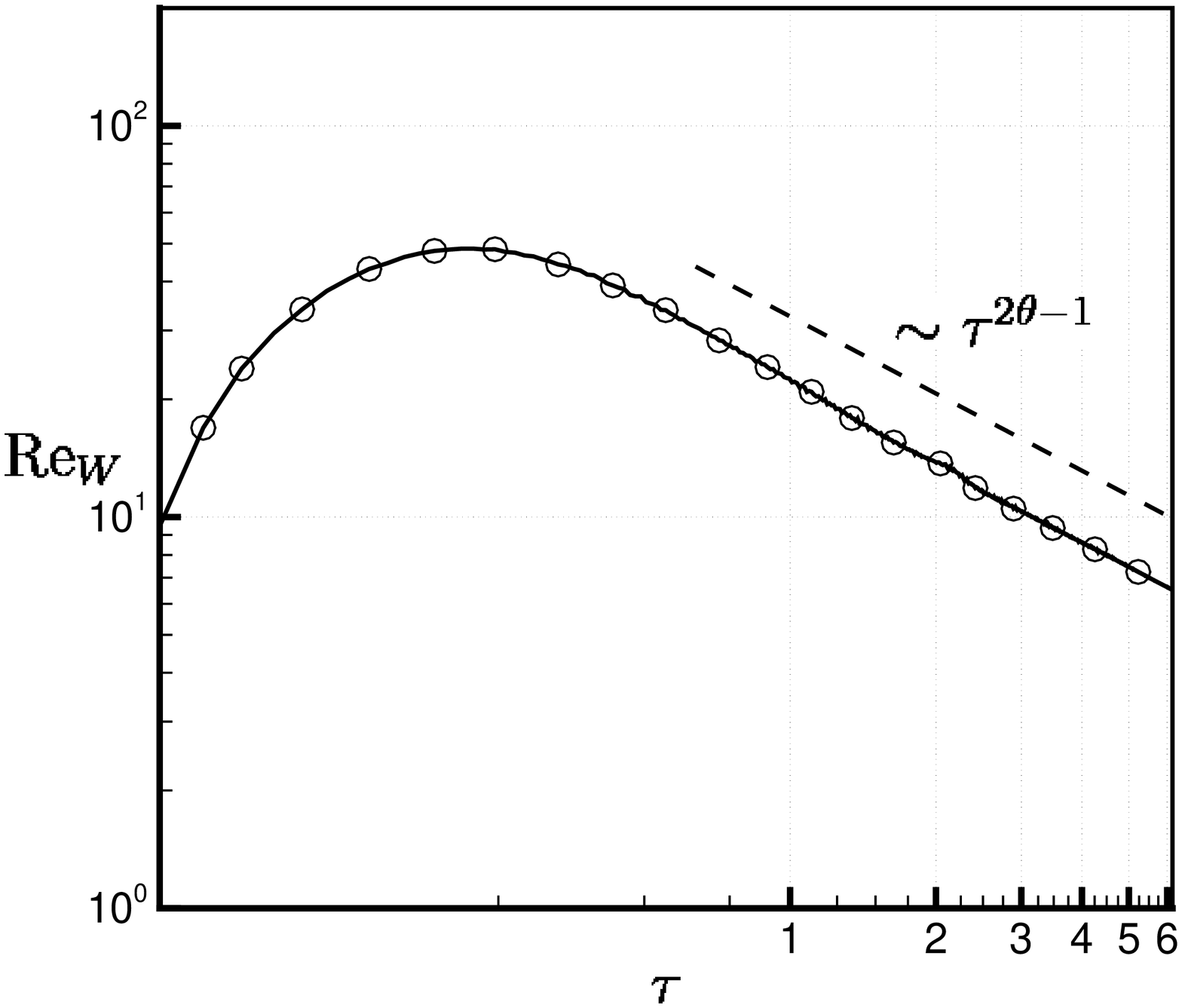}%
		%		\caption{Molecular mixing fraction}%
	\end{subfigure}	
	\caption{\label{fig:reynolds} Time histories of Reynolds numbers based on Taylor microscale and integral width on the $720\times 512^2$ grid.}
\end{figure}

\textcolor{black}{This section will present a comparison of results between the current DNS and ILES of the same initial condition as presented in the $\theta$-group collaboration} \cite{Thornber2017}. The ILES results on a grid resolution of $720\times 512^2$ (using the same numerical algorithm) are included in the plots below for comparison and are given by dashed grey lines. \textcolor{black}{In order to facilitate comparisons with prior DNS in the literature, in particular prior DNS studies of RMI, the temporal variation in Taylor microscale Reynolds number is given in Fig. \ref{fig:reynolds}. The Reynolds number based on integral width is also presented, showing that it is of similar magnitude. Indeed, at early time both Re$_T$ and Re$_W$ reach a maximum of 49.8 and 48.6 respectively, after which they both decay to final values of 11.3 and 6.5.}

All of the statistics presented here should exhibit some dependence on Reynolds number, at least up until some transitional Reynolds number and necessary time to transition is reached \cite{Dimotakis2000,Zhou2007}. Establishing whether a high Reynolds number limit exists, and whether the results from resolved simulations at lower Reynolds numbers can point to it, is of crucial importance for further understanding the nature of the mixing transition in RMI induced flows. For the initial condition considered here the flow is becoming more well resolved at late time and therefore it is only necessary to use very high grid resolutions at early times to fully capture the flow behaviour. This decreasing resolution requirement at late time can be explained by considering the variation in Reynolds number throughout the simulation. Assuming the mixing layer at late time can be described by a single length scale such as the integral width $W$, then given a Reynolds number for the layer defined as $\textrm{Re}=\dot{W}W/\overline{\nu}$ it is straightforward to show that if the late-time growth obeys a power law $W\propto t^\theta$ then
\begin{equation}
\textrm{Re}\propto \frac{\theta t^{\theta-1}t^\theta}{\nu}=\frac{\theta t^{2\theta-1}}{\nu}
\end{equation}
This implies that the Reynolds number either decreases with time if $\theta < \frac{1}{2}$ or increases with time if $\theta > \frac{1}{2}$. For the current narrowband perturbation the growth rate exponent $\theta$ is 0.2185 (similar to the ILES where $\theta=0.2203$) and hence the layer Reynolds number is expected to decrease at late time, which is indeed the case as shown in Fig. \ref{fig:reynolds}. This is also consistent with the observation of decreasing resolution requirements for DNS as time progresses, which suggests for the possibility of varying the mesh resolution during the simulation so as to achieve a higher Reynolds number while still remaining fully resolved throughout in future computations.

\subsection{Integral Measures}
\label{subsec:mix}
\begin{figure}
	\centering
	\begin{subfigure}[b]{0.48\textwidth}
		\includegraphics[width=\textwidth]{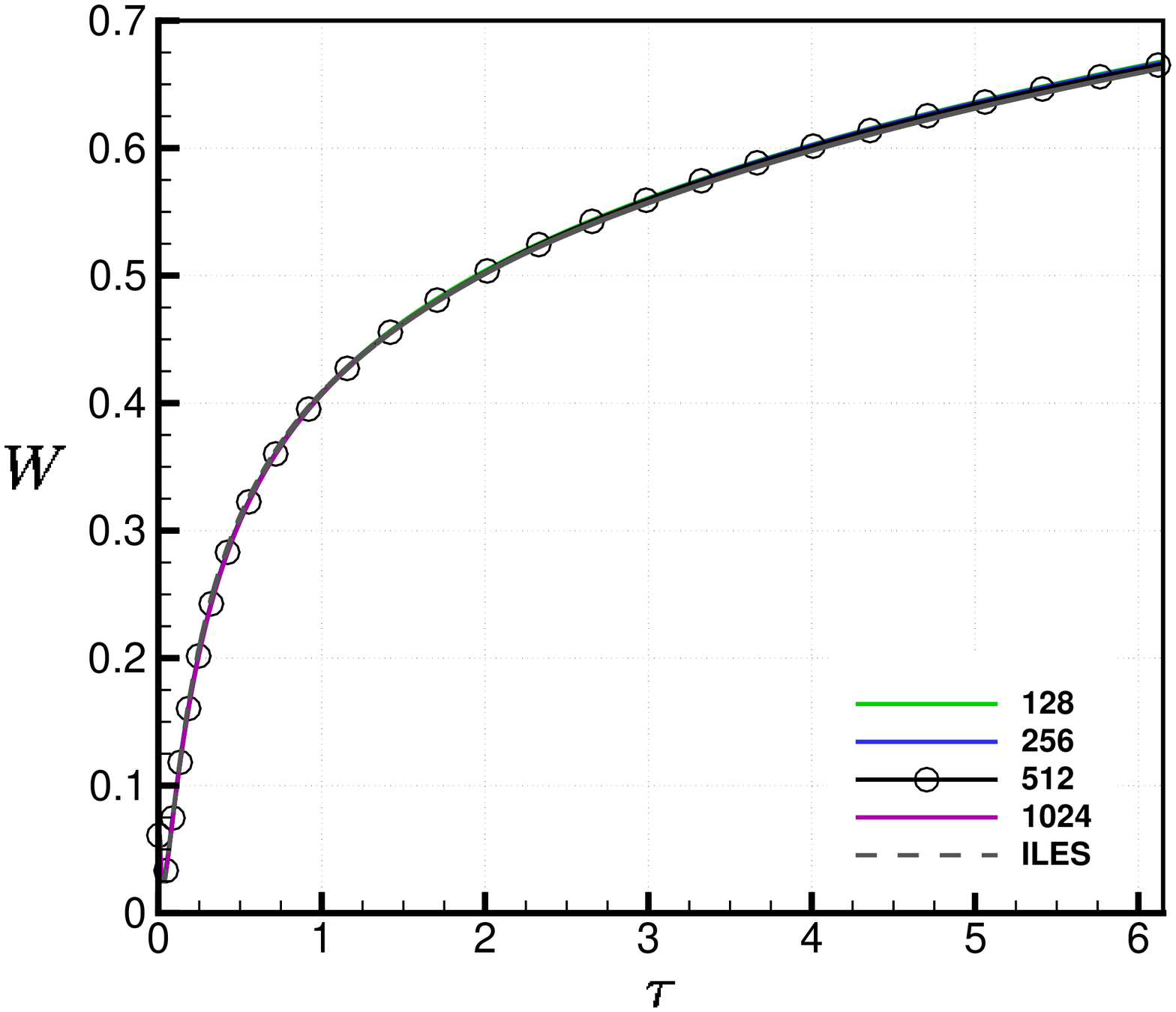}%
		%		\caption{Integral Width}
	\end{subfigure}
	\hfill
	\begin{subfigure}[b]{0.48\textwidth}
		\includegraphics[width=\textwidth]{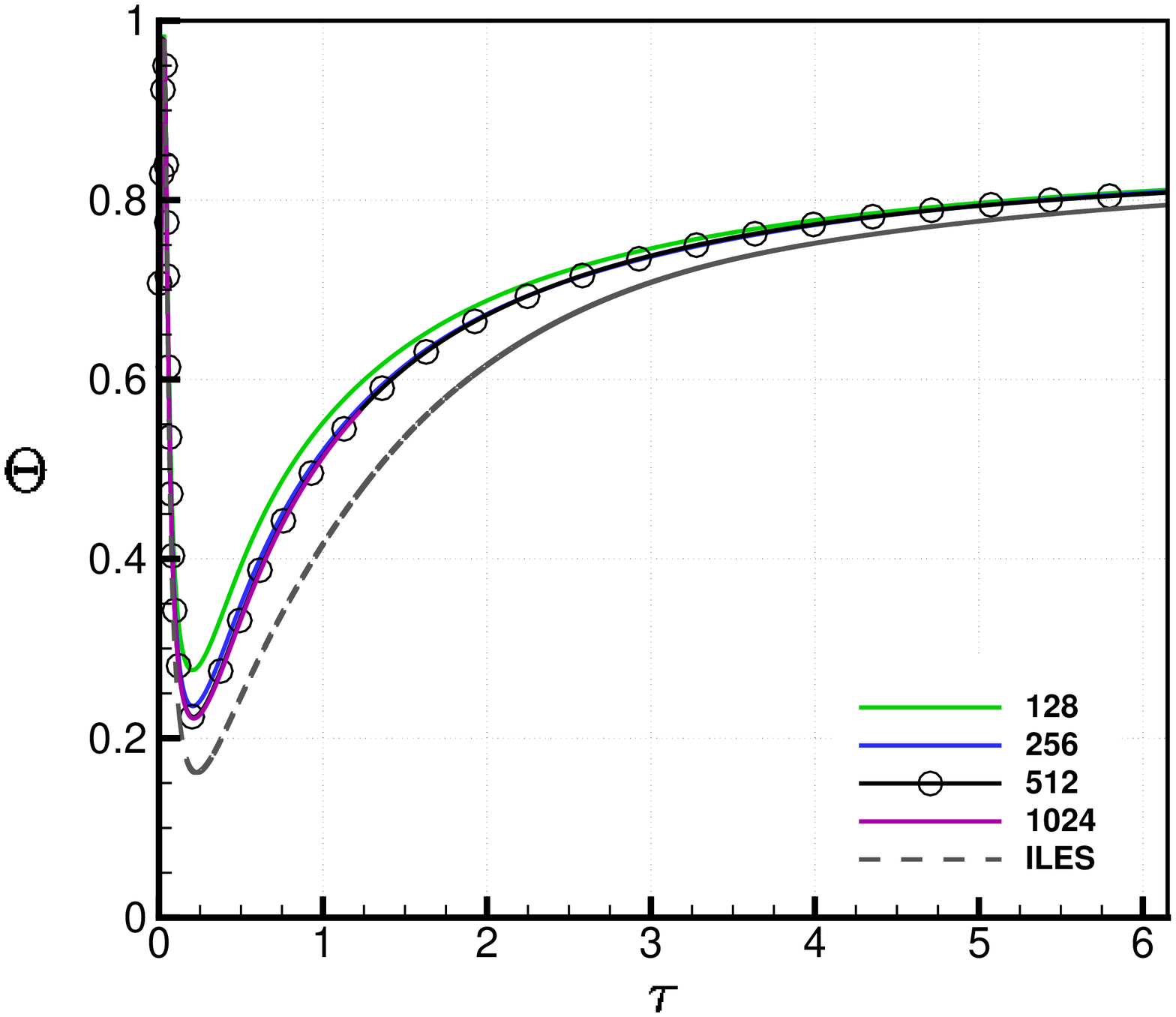}%
		%		\caption{Molecular Mixing Fraction}%
	\end{subfigure}	
	
	\begin{subfigure}[b]{0.48\textwidth}
		\includegraphics[width=\textwidth]{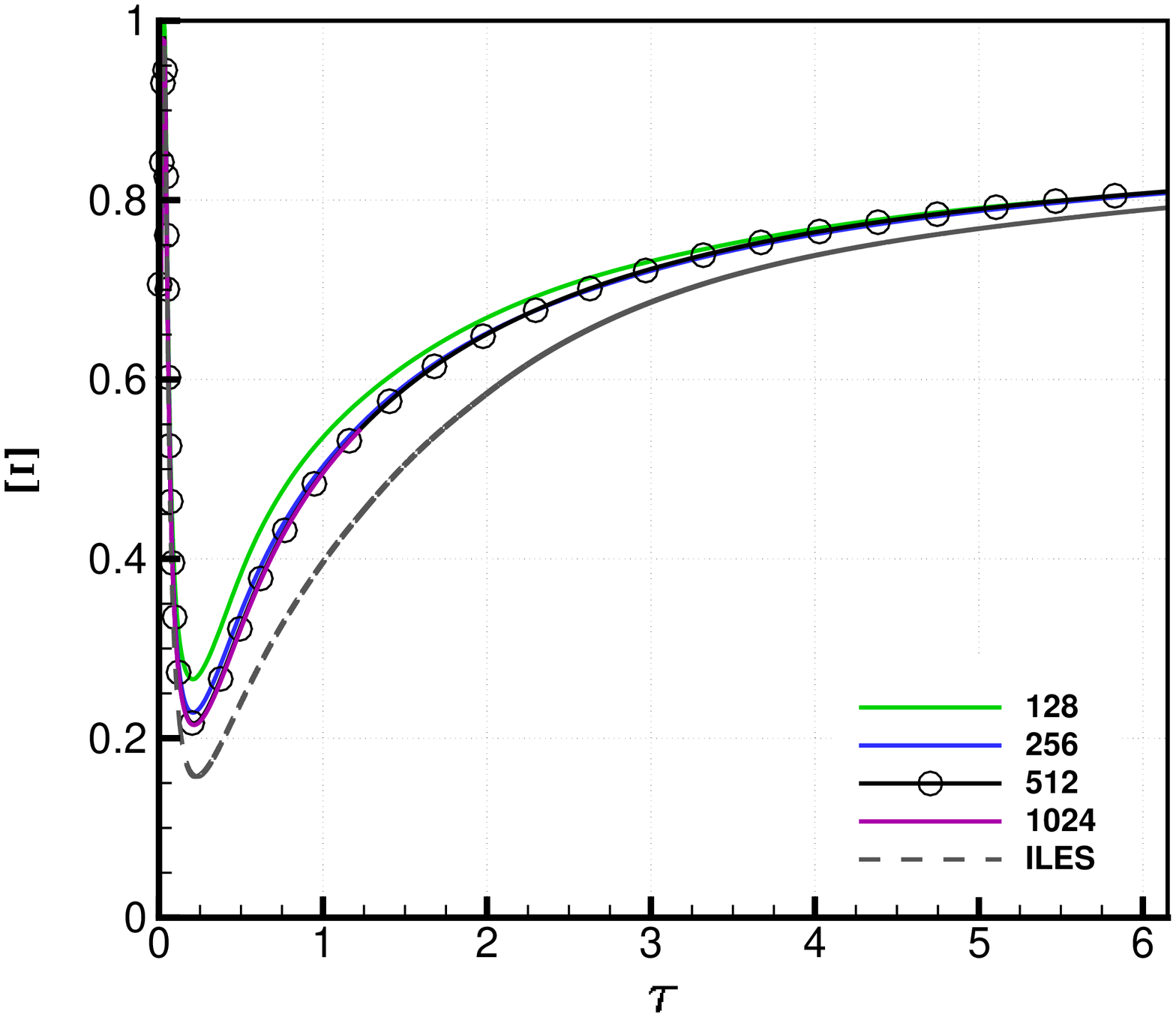}%
		%		\caption{Mixing Parameter}%
	\end{subfigure}	
	\hfill
	\begin{subfigure}[b]{0.48\textwidth}
		\includegraphics[width=\textwidth]{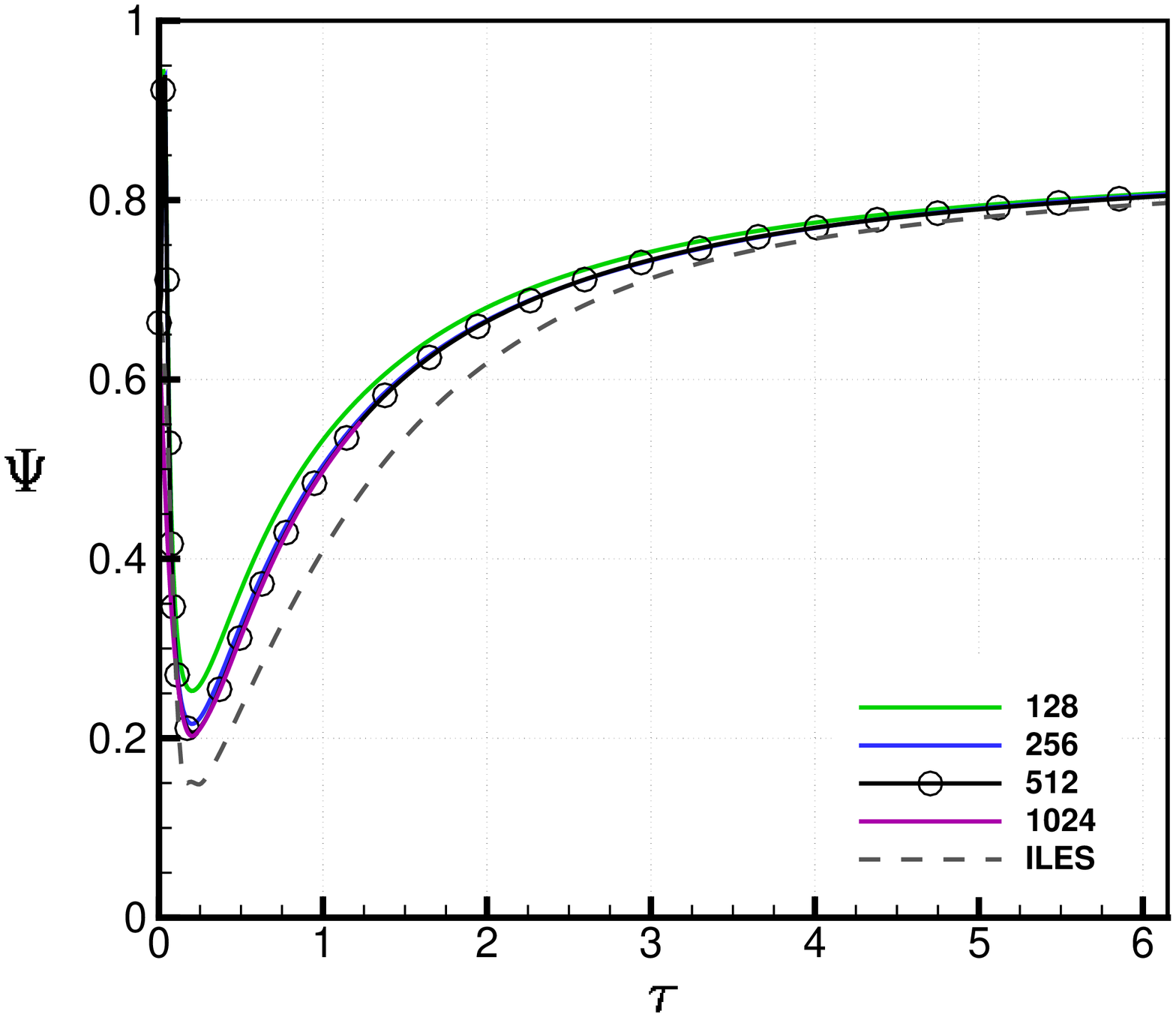}%
		%		\caption{Mixed mass}%
	\end{subfigure}	
	\caption{\label{fig:integral} Time histories of integral width and mixing parameters comparing the DNS and ILES results.}
\end{figure}

The temporal evolution of integral width and the various mixing fractions is given in Fig. \ref{fig:integral}. There is little difference in the integral width predicted by the current DNS versus that of the ILES, particularly at early time. This indicates that the largest scales are still evolving mostly independently of the dissipation mechanism in the present DNS at this given Reynolds number. This trend is expected to eventually no longer hold as the Reynolds number is further decreased \cite{Walchli2017}. The end time (non-dimensional) integral width is 0.6676 for the DNS compared to 0.6628 for the ILES, a difference of 0.72\%. \textcolor{black}{This corresponds to a (non-dimensional) visual width of $\delta=5.28$ and $\delta=5.24$ respectively, where $\delta$ is defined as the distance between $x$ locations where $\langle f_1 \rangle=0.99$ and $\langle f_1 \rangle=0.01$.} The end time values of the mixing fractions $\Theta$, $\Xi$ and $\Psi$ are 0.8064, 0.8073 and 0.8048 respectively for the DNS, compared to 0.7945, 0.7915 and 0.7968 for the ILES. This equates to an average difference of 1.48\% across all three metrics. \textcolor{black}{Note that at the initial time, $\Theta$, $\Xi$ and $\Psi$ are not equal to 1 since the layer does not begin in a purely homogeneous state due to the initial perturbation and is therefore not perfectly mixed (i.e. $\langle f_1 f_2 \rangle \ne \langle f_1 \rangle \langle f_2 \rangle$).} The slightly larger values of $W$, $\Theta$, $\Xi$ and $\Psi$ for the DNS compared to the ILES are caused by extra spreading of the layer due to molecular diffusion at early time and inhibition of turbulence. It is also interesting to note the degree of similarity of the late time values of the mixing fractions  between the DNS and ILES (in particular the agreement in $\Psi$ at late time), despite the inhibition of fine-scale turbulent motions in the DNS. This indicates that at late time the mixing is dominated by large scale motions, which are very similar for the the DNS and ILES.

\begin{figure}
	\centering
	\begin{subfigure}[b]{0.49\textwidth}
		\includegraphics[width=\textwidth]{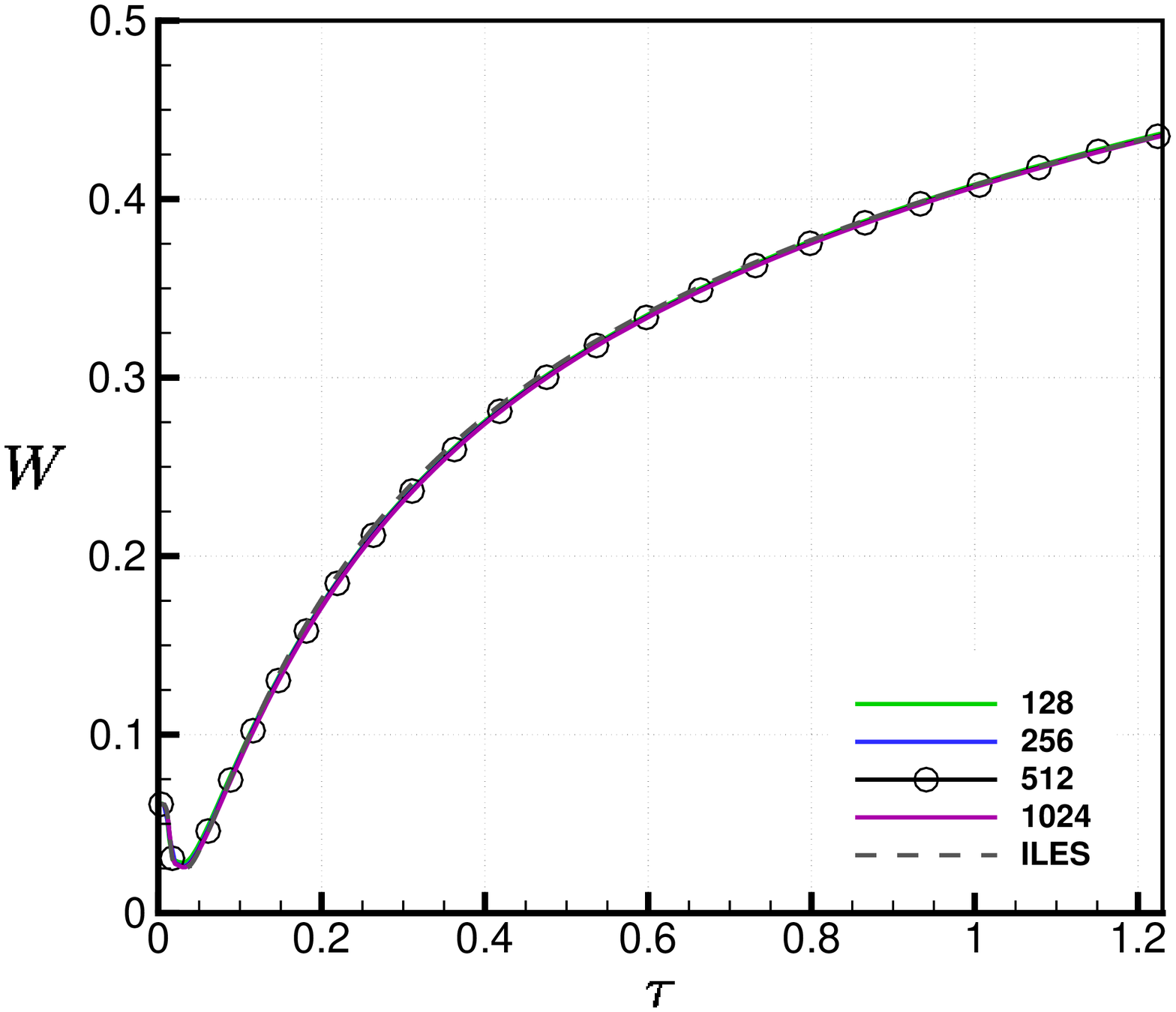}%
		%		\caption{Integral width}%
	\end{subfigure}
	\hfill
	\begin{subfigure}[b]{0.49\textwidth}
		\includegraphics[width=\textwidth]{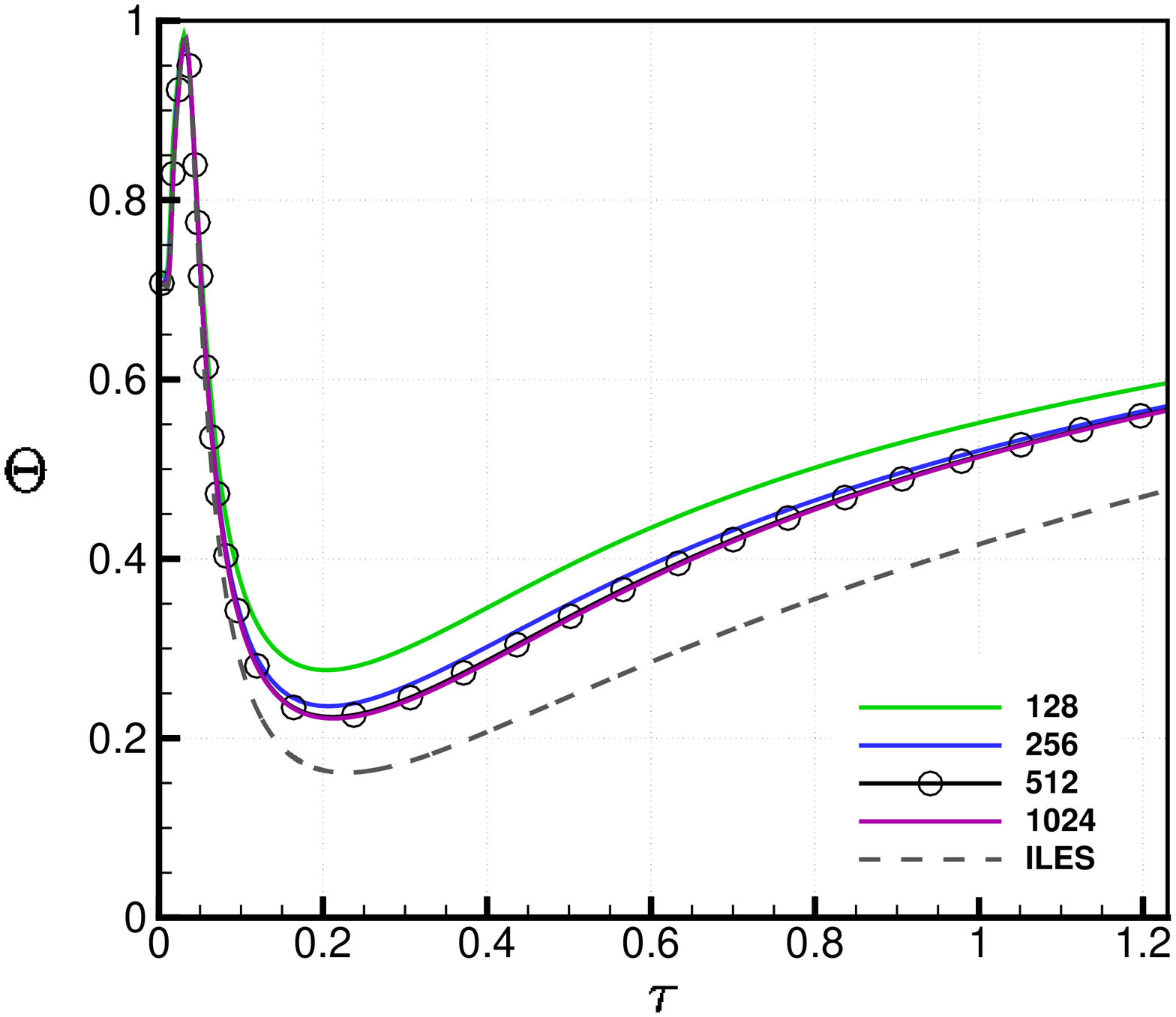}%
		%		\caption{Molecular mixing fraction}%
	\end{subfigure}	
	\caption{ Integral width and molecular mixing fraction at early time.}
	\label{fig:integralearly}
\end{figure}

Examining the early time behaviour of $W$ and $\Theta$ in Fig. \ref{fig:integralearly} shows the initial compression of the layer due to the shock, which corresponds to a minimum in $W$ and a maximum in $\Theta$. At a slightly later time $\Theta$ obtains a minimum value, which corresponds to when the interface is most highly stretched. For the DNS, this stretching of the interface due to the initial impulse is balanced by molecular diffusion due to gradients across the interface as well as the onset of any secondary instabilities. As a result, the value and temporal location of minimum mix is conjectured to be a function of the initial Reynolds number. The values of $\Theta$ and $\Xi$ at this point are 0.2221 and 0.2148 respectively, both occurring at a non-dimensional time of $\tau=0.2126$, which is earlier than the time of maximum scalar dissipation rate. For the ILES, the stretching due to the initial impulse is balanced purely by secondary instabilities and the implicit sub-grid model, with the minimum in $\Theta$ and $\Xi$  of 0.1616 and 0.1572 occurring slightly later at $\tau=0.2274$ and $\tau=0.2268$ respectively. Whether this is an accurate estimate of the high Reynolds number limit (if such a limit exists) is an open question since at this time the evolution of the large scales may not be independent of the dissipation mechanism. At the very least this value represents the inviscid limit for a given grid resolution and algorithm.

\subsection{Turbulent Kinetic Energy}
\label{subsec:tke}
\begin{figure}
	\centering
	\begin{subfigure}[b]{0.48\textwidth}
		\includegraphics[width=\textwidth]{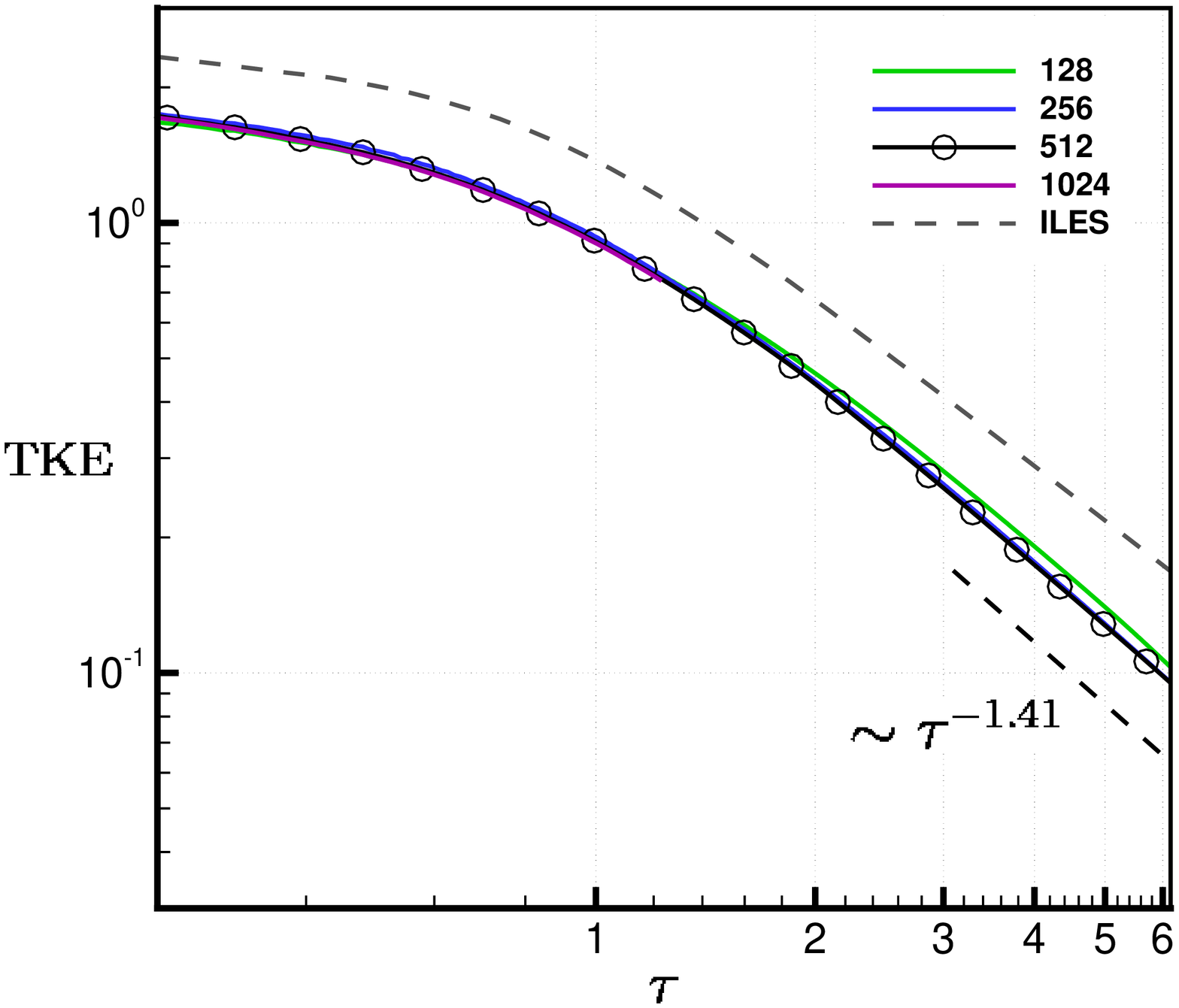}%
		%    	\caption{Turbulent Kinetic Energy}%
	\end{subfigure}	
	\hfill
	\begin{subfigure}[b]{0.48\textwidth}
		\includegraphics[width=\textwidth]{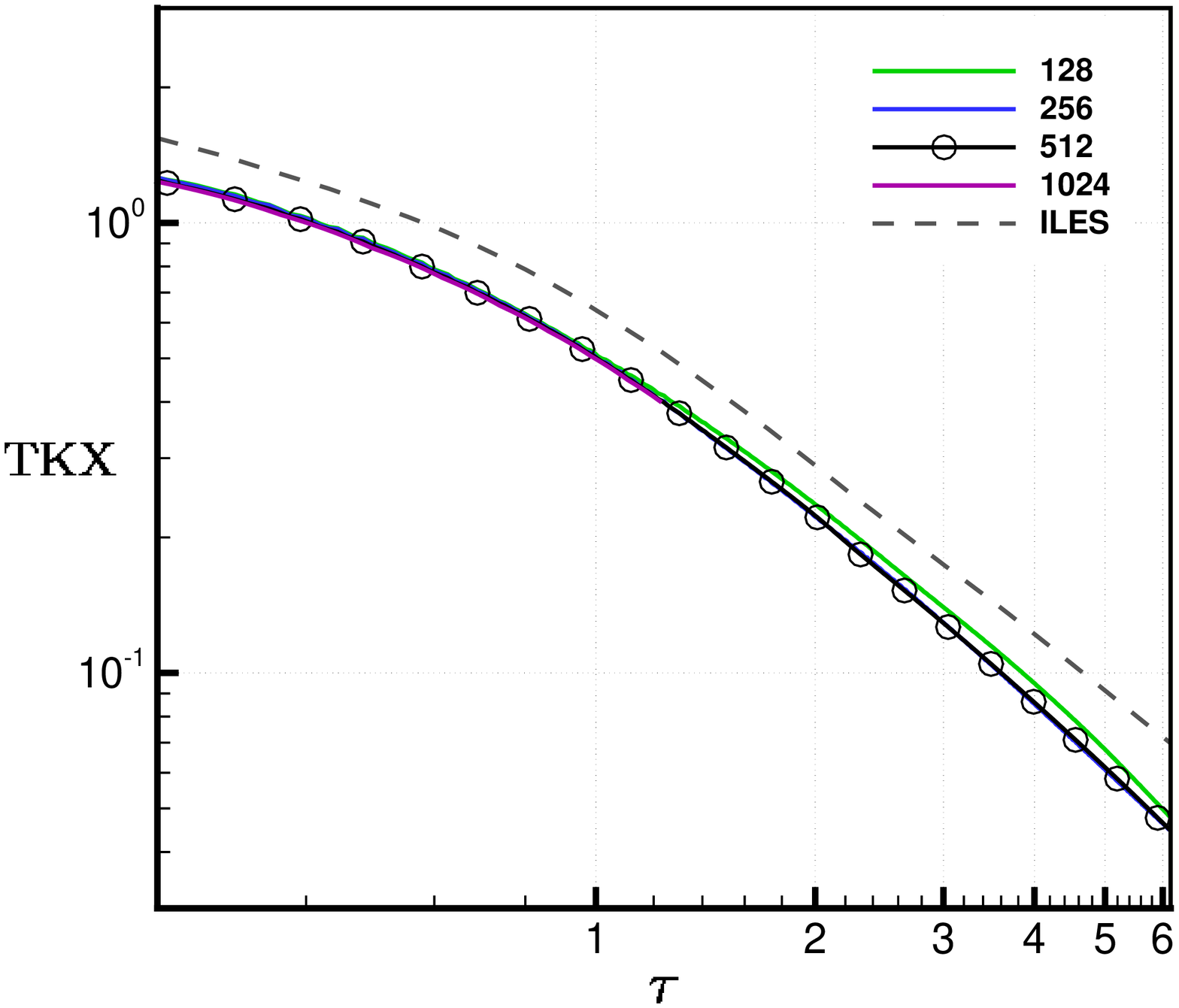}%
		%		\caption{TKX}%
	\end{subfigure}	
	
	\begin{subfigure}[b]{0.48\textwidth}
		\includegraphics[width=\textwidth]{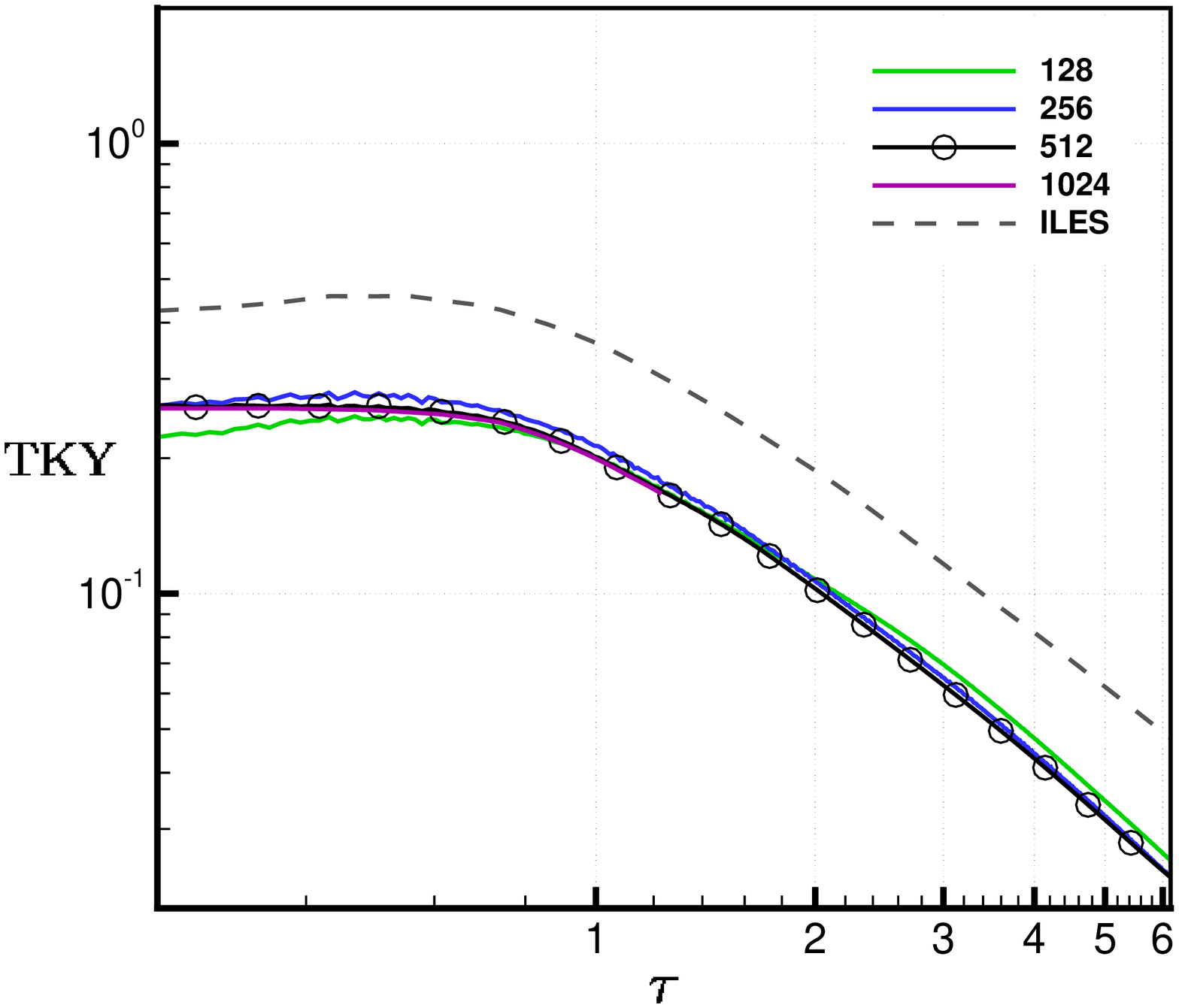}%
		%		\caption{TKY}%
	\end{subfigure}	
	\hfill
	\begin{subfigure}[b]{0.48\textwidth}
		\includegraphics[width=\textwidth]{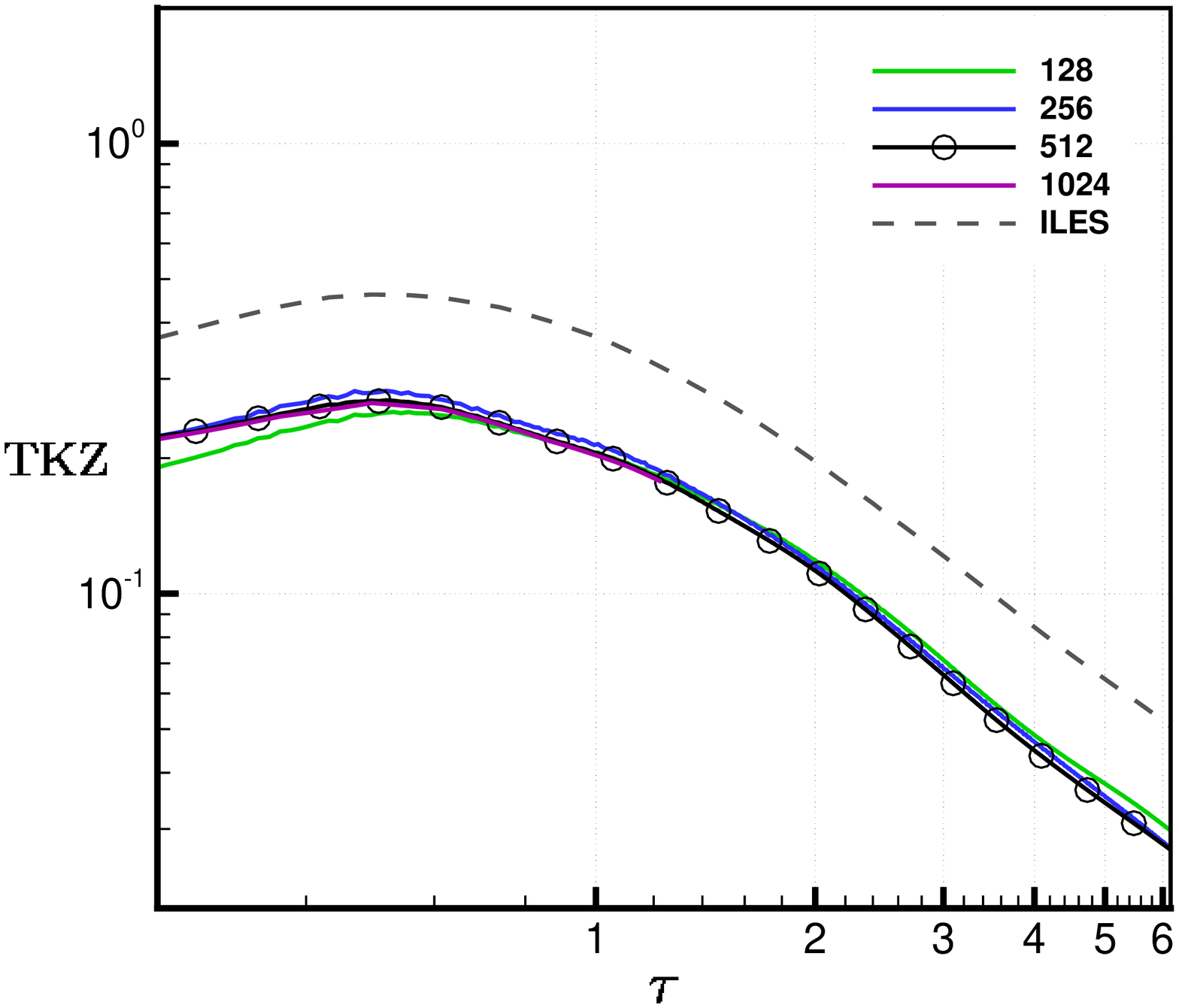}%
		%    	\caption{TKZ}%
	\end{subfigure}	
	\caption{\label{fig:tke} Time histories of turbulent kinetic energy components comparing the DNS and ILES results.}
\end{figure}

\textcolor{black}{Turbulent kinetic energy may be compared between the DNS and ILES since in the high Reynolds limit it depends only on the large scales and as such will converge in a similar manner to the integral width.} Comparing the temporal evolution of turbulent kinetic energy (TKE) in Fig. \ref{fig:tke}, the DNS results show a higher decay rate of $\tau^{-1.41}$, compared to a rate of $\tau^{-1.25}$ for the ILES. The decay rate of $-1.41$ for the DNS is very close to the theoretical value of $-10/7$ for homogeneous decaying turbulence (assuming a $k^4$ Batchelor form for the large scales \cite{Thornber2016}), however this is likely just a coincidence given that the decay rate of TKE does not match the value that can be calculated from the observed growth of the integral width assuming self-similarity \cite{Thornber2010}. Nor does it correspond with the observed decay rate of enstrophy as shown in Fig. \ref{fig:dissipation}, which should scale the dissipation rate. In addition, the low Reynolds number, short time scale and anisotropy of the mixing layer (see below) indicate that instead it is far more likely that the increased decay rate is simply due to the additional dissipation and a lack of scale separation. 

\begin{figure}
	\centering
	\begin{subfigure}[b]{0.49\textwidth}
		\includegraphics[width=\textwidth]{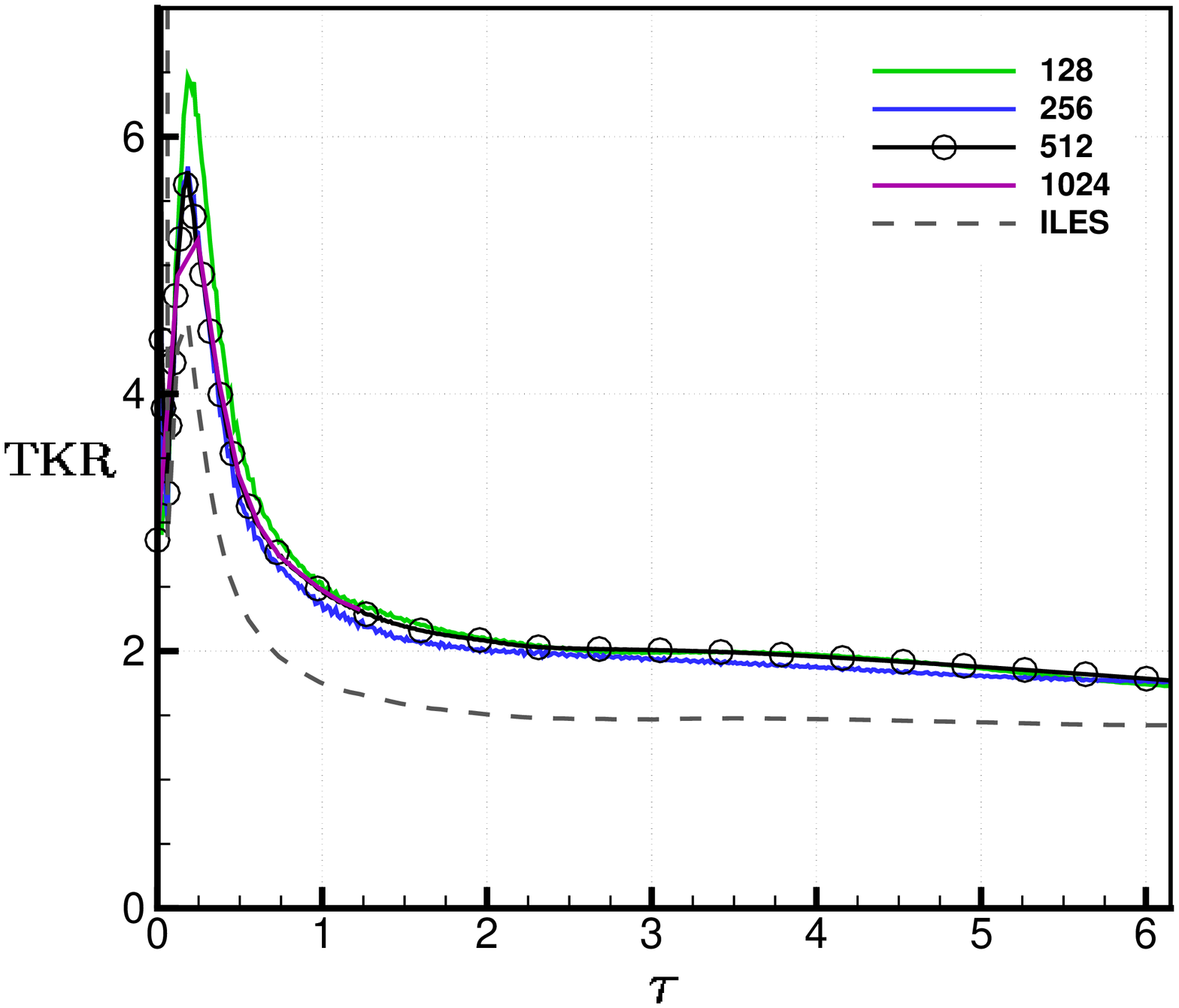}%
		%		\caption{Integral width}%
	\end{subfigure}
	\hfill
	\begin{subfigure}[b]{0.49\textwidth}
		\includegraphics[width=\textwidth]{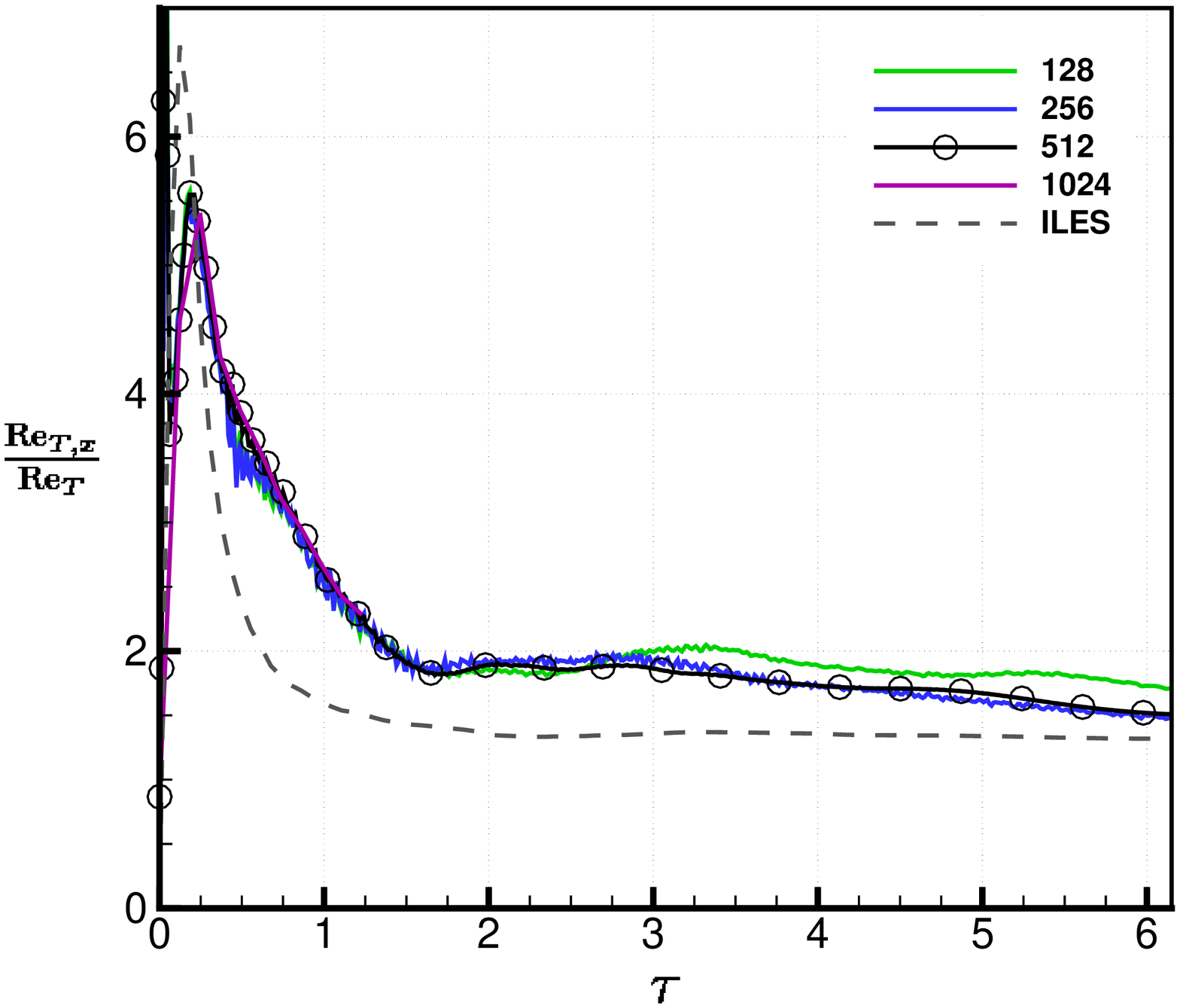}%
		%		\caption{Molecular mixing fraction}%
	\end{subfigure}	
	\caption{\label{fig:anisotropy} Time histories of the ratios of TKE components and Taylor microscale Reynolds numbers.}
\end{figure}

Fig. \ref{fig:tke} also contains the evolution of the individual components of TKE. All components of TKE are reduced in magnitude compared to the ILES results, however the $x$-component of turbulent kinetic energy has a smaller reduction than in the $y$ and $z$ directions, indicating that there is an increase in anisotropy with increased viscous dissipation. This suggests that viscosity has suppressed secondary instabilities which transfer kinetic energy from the $x$ direction to the $y$ and $z$ directions. \textcolor{black}{This reduced transfer of energy to the perpendicular directions in the DNS can be quantified by considering the ratio of turbulent kinetic energy components $\mathrm{TKR}=(2\times \mathrm{TKX})/(\mathrm{TKY}+\mathrm{TKZ})$ as well as the ratio of Taylor microscale Reynolds numbers $\textrm{Re}_{T,x}/\textrm{Re}_T$. Fig. \ref{fig:anisotropy} shows the temporal evolution of both of these ratios for the DNS and ILES (note that although $\textrm{Re}_T$ cannot be defined for the ILES, the ratio is a valid quantity as $\langle \nu \rangle$ cancels).} A peak in these ratios (ignoring the initial compression) occurs at a non-dimensional time of $\tau=0.1845$, slightly before the observed minimum in mixing parameters. At the latest time, the ratio of turbulent kinetic energy components is 1.692 for the DNS compared to 1.423 for the ILES, while the ratio of Taylor microscale Reynolds numbers is 1.508 and 1.318 for the DNS and ILES respectively. \textcolor{black}{Thus both metrics show that there is still a significant amount of anisotropy in the flow at the latest time considered. Although this anisotropy is still decreasing (with the DNS results also approaching those of the ILES), an analysis of the quarter-scale case data from the $\theta$-group collaboration shows that persistent anisotropy still remains at much later non-dimensional times \cite{GroomAFMC,Thornber2017}, in good agreement with the theoretical results of Soulard \textit{et al.} \cite{Soulard2018}.}

\begin{figure}
	\centering
	\begin{subfigure}[b]{0.48\textwidth}
		\includegraphics[width=\textwidth]{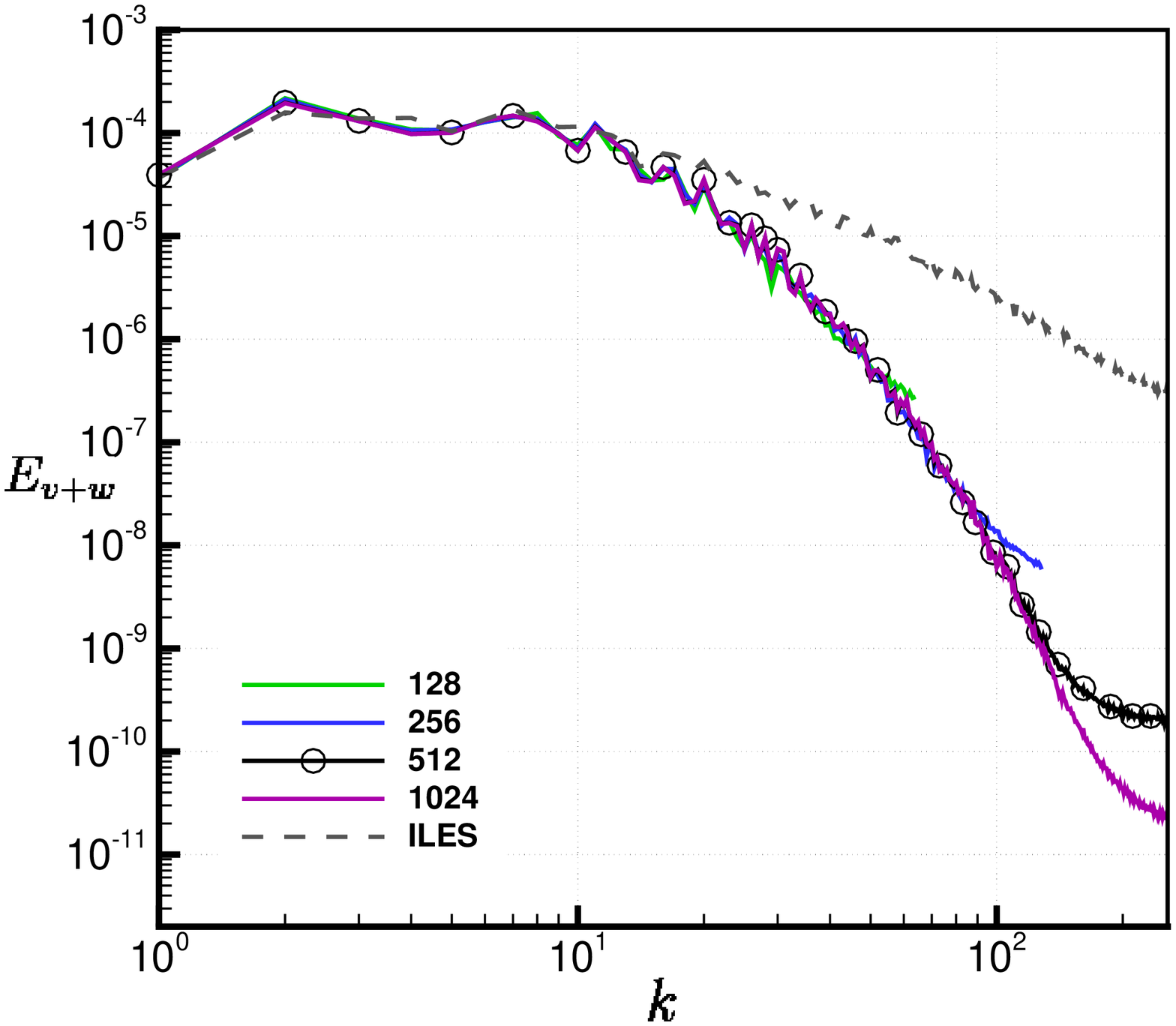}%
%		\subcaption{$\tau=1.23$}%
	\end{subfigure}
	\hfill
	\begin{subfigure}[b]{0.48\textwidth}
		\includegraphics[width=\textwidth]{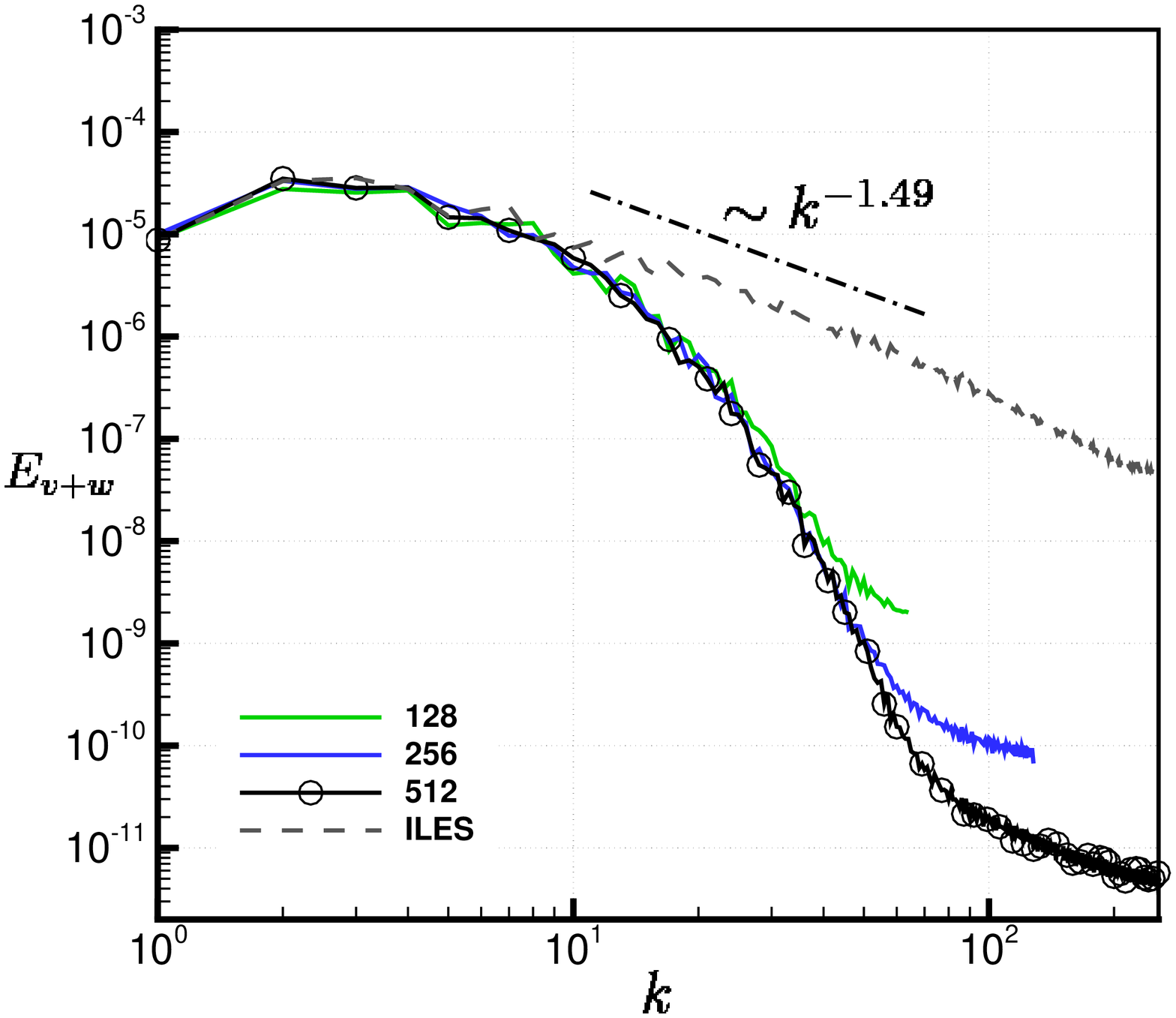}%
%		\subcaption{$\tau=6.15$}%
	\end{subfigure}	
	\begin{subfigure}[b]{0.48\textwidth}
	\includegraphics[width=\textwidth]{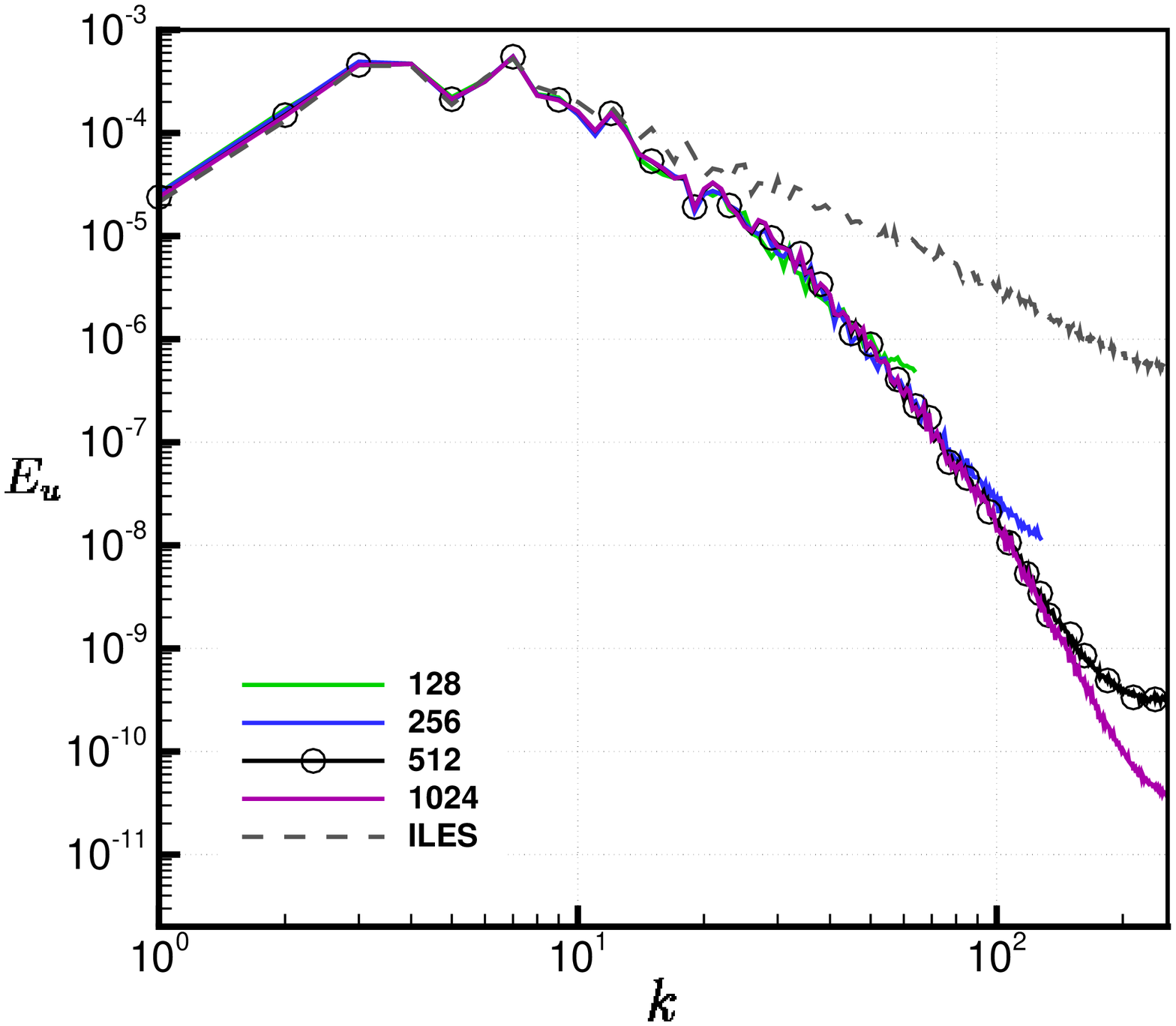}%
	\subcaption{$\tau=1.23$}%
\end{subfigure}
	\hfill
	\begin{subfigure}[b]{0.48\textwidth}
		\includegraphics[width=\textwidth]{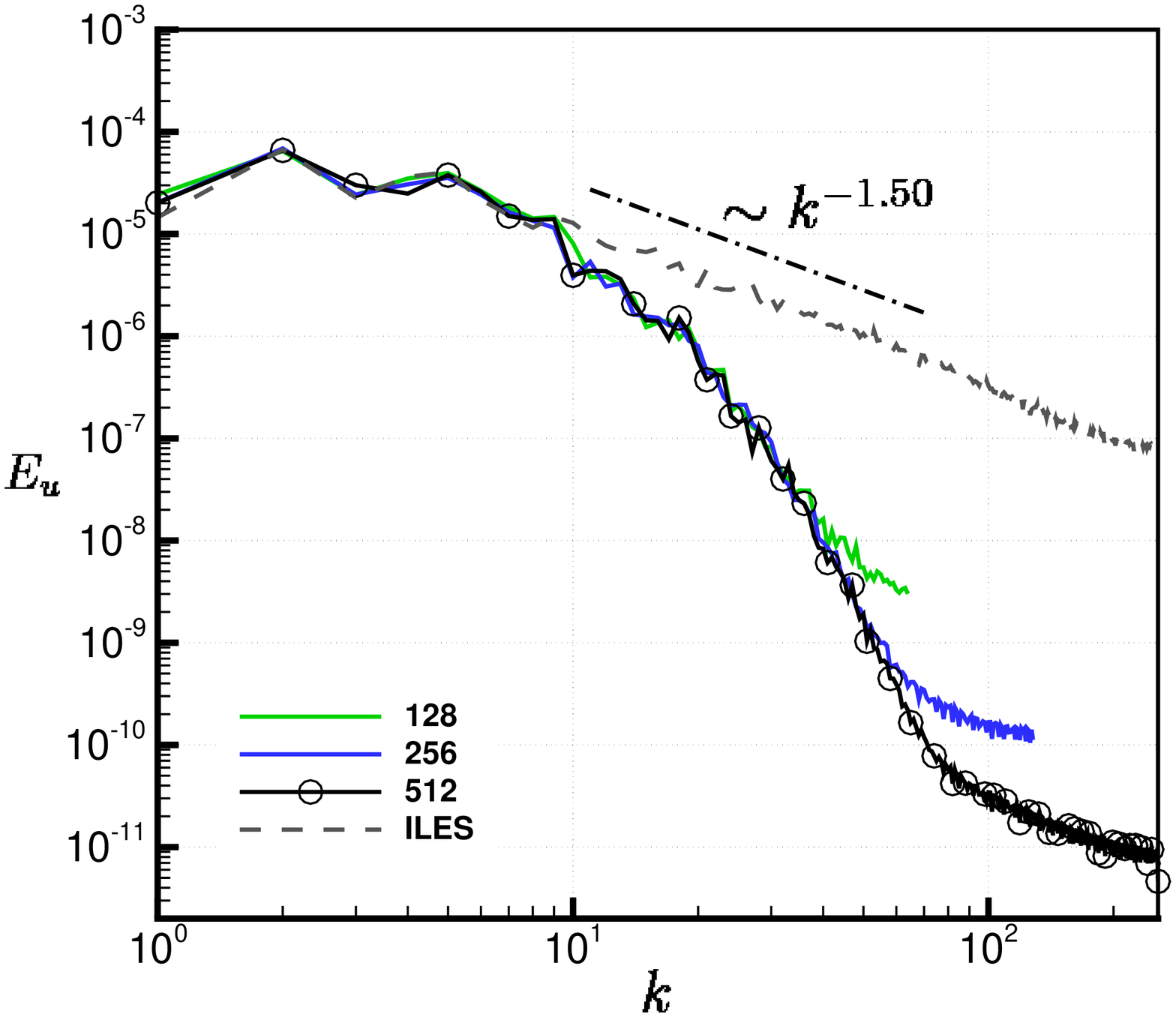}%
		\subcaption{$\tau=6.15$}%
	\end{subfigure}	
	\caption{\label{fig:spectra1} Power spectra of turbulent kinetic energy, calculated at the mixing layer centre plane.}
\end{figure}

The wavenumber dependence of turbulent kinetic energy is examined by computing the 2D radial power spectra, shown in Fig. \ref{fig:spectra1}. \textcolor{black}{Power spectra of both the in-plane ($E_{v+w}$) and the out-of-plane ($E_u$) components are shown.} The turbulent kinetic energy spectra for the DNS show no signs of an inertial subrange being present, indicating that a significant separation of scales is not present. When comparing with the ILES results it is clear that the reduction in TKE is due to viscous suppression of the higher wavenumbers; the large scales are in good agreement. \textcolor{black}{In particular, this agreement is stronger for the $E_u$ spectra than the $E_{v+w}$ spectra, again showing that the suppression of secondary instabilities reduces the transfer of energy from the $x$ direction to the $y$ and $z$ directions in accordance with Fig. \ref{fig:tke}. This is also the mechanism behind the increased levels of anisotropy observed in the DNS compared to the ILES.}

The excellent agreement in the large scales between the DNS and ILES gives support to the assumption that the growth of the integral length scale is independent of the mechanism of dissipation \cite{Thornber2017}, one of the key tenets of the ILES philosophy. A $k^{-3/2}$ inertial range is present in the ILES data across at least half a decade, consistent with the theory of Zhou \cite{Zhou2001} and previous simulations \cite{Thornber2010,Thornber2016}. It is anticipated that as the Reynolds number increases the agreement between the DNS and ILES results will extend to higher and higher wavenumbers, ultimately resulting in the development of an inertial subrange once some critical Reynolds number is reached. This will be the focus of future work. The spectra also show that the Reynolds number is decreasing at late-time as the energy contained at high wavenumbers is decreasing, in line with the argument made in Section \ref{sec:results}.

\section{Conclusion}
\label{sec:conclusion}
The narrowband multimode Richtmyer-Meshkov instability has been investigated in a planar geometry using direct numerical simulation of a well-defined deterministic initial condition. A grid converged solution was clearly demonstrated by considering the temporal evolution of enstrophy and scalar dissipation rate as well as their power spectra. Satisfactory convergence in these quantities was also shown to be consistent with the Kolmogorov scale being resolved on the finest grid for the entire duration of the simulation. By comparing the results with those of an implicit large eddy simulation of the same initial condition a detailed account of the early time transitional behaviour, and insight into the high Reynolds number limit, can be made as follows. After compression of the interface by the shock wave at $\tau\approx 0.01$, a minimum in the integral width $W$ and maximum in the mixing fractions $\Theta$ and $\Xi$ occurs at $\tau=0.0300$ corresponding to the point of peak compression. The anisotropy of the layer, as measured by the ratios of components of turbulent kinetic energy as well as directional Taylor microscale Reynolds numbers, has an initial peak at the time of compression before declining and then peaking again at a time of $\tau=0.1845$ in both the DNS and ILES. This is followed shortly thereafter by a minimum in $\Theta$ and $\Xi$, at time $\tau=0.2126$ in the DNS and at  $\tau=0.2268$ and $\tau=0.2274$ in the ILES. The enstrophy in the DNS decays monotonically throughout the simulation after the shock interaction, while the scalar dissipation rate exhibits a maximum at $\tau=0.4931$, beyond which it also decays monotonically. The maximum in scalar dissipation rate is indicative of the time when energy coupling to the highest modes has occurred and the flow is becoming damped at these scales. Late time decay rates for turbulent kinetic energy, enstrophy and scalar dissipation rate were calculated, with the TKE found to be decaying at a faster rate at late time in the DNS than in the ILES. Examination of the power spectra at various points in time showed that this increased decay is confined to the high wavenumber portion of the spectrum, at low wavenumbers there is good agreement between the DNS and ILES results (prior to the start of the inertial range in the ILES spectrum).

Given these results it can be expected that as the initial Reynolds number increases, the point of minimum mix will decrease and occur later in time, the peak in enstrophy and scalar dissipation rate will increase and occur later in time, the late time decay rates of TKE and dissipative quantities will decrease and the layer will become less anisotropic as more kinetic energy is transferred to the transverse directions. Of key interest is how rapidly these changes will occur as a function of Reynolds number, and whether there is some critical Reynolds number beyond which the dependence on Reynolds number is quickly lost. This will be investigated further in future work with multiple direct numerical simulations across a range of Reynolds numbers. If such an understanding can be established, then along with the LES computations that provide the physics of the self-similar layer, a complete picture of the planar narrowband RMI will be obtained from initiation and transition through to self-similarity. This will then allow for the development of models that better capture the behaviour of the mixing layer prior to the development of a fully self-similar state.

\section{Acknowledgements}
\label{sec:acknowledgements}
This research was supported under Australian Research Council's Discovery Projects funding scheme (project number DP150101059). The authors would like to acknowledge the computational resources at the National Computational Infrastructure through the National Computational Merit Allocation Scheme, as well as the Sydney Informatics Hub and the University of Sydney's high performance computing cluster Artemis, which were employed for all cases presented here. The authors also wish to acknowledge Prof. David Youngs for helpful discussions regarding the late-time behaviour of narrowband RMI.
%\end{linenumbers}
%\appendix

%\section{Additional quantities}
%\label{}

  \bibliographystyle{elsarticle-num} 
  \bibliography{bibliography}


%\begin{thebibliography}{00}


%\bibitem{}
%
%\end{thebibliography}
\end{document}